\documentclass[12pt,a4paper]{article}
\pdfoutput=1
\usepackage{graphicx}
\usepackage[T1]{fontenc}
\usepackage[sc,osf]{mathpazo}
\usepackage{a4wide}  
\usepackage{latexsym,amsthm,amsfonts,amsmath,mathrsfs,amssymb}
\usepackage[unicode,implicit]{hyperref}
\hypersetup{%
  pdftitle    = {On a family of alpha prime-corrected solutions of the Heterotic
      Superstring effective action}
  pdfkeywords = {strings, superstrings, supergravity, heterotic, corrections,
    alpha prime, instantons, solitons, dyons},
  pdfauthor   = {Samuele Chimento, Patrick Meessen, Tomas Ortin, Pedro
    F. Ramirez and Alejandro Ruiperez},
  plainpages  = true,
  colorlinks  = true,
  citecolor   = blue,
  urlcolor    = red,
  linkcolor   = black
}
\newcommand{\hepth}[1]{{\tt
\href{http://www.arXiv.org/abs/hep-th/#1}{hep-th/#1}}}

\newcommand{\arxiv}[1]{{\tt arXiv:\href{http://www.arXiv.org/abs/#1}{#1}}}
\newcommand{\doi}[1]{{\tt DOI:\href{http://dx.doi.org/#1}{#1}}}

\usepackage{tikz}\newcommand{\FPAUO}[2]{
\tikz[scale=.13,
         Uniovi/.style={color=green!51!blue, fill=green!51!blue}
 ] {
 \fill[Uniovi] (0,0) circle (10);
 \fill[white] (0,7) circle (1.5);
 \draw[Uniovi] (-2,7.5) rectangle (2,5.5);
 \fill[white] (-0.3,6.6) rectangle (0.3,0);   % 1.7 cm 
 \fill[white] ( -0.9,6.2) rectangle (.9 ,5.6);
 \fill[white] (-1.4, 5.2) rectangle (1.4, 4.6);
 \fill[white] (0,0) ellipse (3.5 and 4);
 \fill[Uniovi] (-2.5,0.3) rectangle (2.5,-0.3);
 \fill[Uniovi] (-2,2.3) rectangle (2,1.7);
 \fill[Uniovi] (-2,-2.3) rectangle (2,-1.7);
 \fill[white] (-4.5,5.5) rectangle (-2.7,4.9);
 \fill[white] (-3.9,6.1) rectangle (-3.3,4.3);
 \fill[white] (4.5,5.5) rectangle (2.7,4.9);
 \fill[white] (3.9,6.1) rectangle (3.3,4.3);
 \foreach \x in { 0,..., 3 }
   \foreach \y in { 0,...,\x}
    {
     \fill[white] (-6-\x*0.7+\y*1.4,3.5-\x *1.97) -- (-5.6-\x*0.7+\y*1.4,2.4-\x *1.97) -- (-6.4-\x*0.7+\y*1.4,2.4-\x *1.97) -- cycle;
     \fill[white] (6-\x*0.7+\y*1.4,3.5-\x *1.97) -- (5.6-\x*0.7+\y*1.4,2.4-\x *1.97) -- (6.4-\x*0.7+\y*1.4,2.4-\x *1.97) -- cycle;
   };
 \draw (0,-6) node[
                               text centered, 
                               color=white, 
                               font={\fontsize{8}{4}\sffamily\selectfont}
                             ] {FPAUO-#1/#2};
}} 
\usepackage{pgfplots}    % This is needed for plotting data
\makeatletter
\@addtoreset{equation}{section}
\makeatother

\begin{document}

\begin{flushright}
\small
\FPAUO{18}{03}\\
IFT-UAM/CSIC-18-018\\
IFUM-1058-FT\\
\texttt{arXiv:1803.04463 [hep-th]}\\
March 11\textsuperscript{th}, 2018\\
\normalsize
\end{flushright}

\begin{center}
 
{\Large {\bf On a family of $\alpha'$-corrected solutions of the\\
 Heterotic Superstring effective action}}%\\[.5cm]
 
\vspace{.5cm}

\renewcommand{\thefootnote}{\alph{footnote}}
{\sl Samuele Chimento$^{1}$}${}^{,}$\footnote{E-mail: {\tt Samuele.Chimento [at] csic.es}},
{\sl Patrick Meessen$^{2}$}${}^{,}$\footnote{E-mail: {\tt meessenpatrick [at] uniovi.es}},
{\sl Tom\'{a}s Ort\'{\i}n$^{1}$}${}^{,}$\footnote{E-mail: {\tt Tomas.Ortin [at] csic.es}},
{\sl Pedro F.~Ram\'{\i}rez$^{3}$}${}^{,}$\footnote{E-mail:  {\tt ramirez.pedro  [at] mi.infn.it}}
{\sl and~Alejandro~Ruip\'erez$^{1}$}${}^{,}$\footnote{E-mail: {\tt alejandro.ruiperez [at] uam.es}},

\setcounter{footnote}{0}
\renewcommand{\thefootnote}{\arabic{footnote}}

\vspace{.3cm}
{\small
{\it Instituto de F\'{\i}sica Te\'orica UAM/CSIC\\
C/ Nicol\'as Cabrera, 13--15,  C.U.~Cantoblanco, E-28049 Madrid, Spain}\\ 
\vspace{0.2cm}

${}^{2}${\it HEP Theory Group, Departamento de F\'{\i}sica, Universidad de Oviedo\\
  Avda.~Calvo Sotelo s/n, E-33007 Oviedo, Spain}\\

\vspace{0.2cm}

${}^{3}${\it 
INFN, Sezione di Milano, Via Celoria 16, 20133 Milano, Italy.
}\\
}

\vspace{.5cm}

%%%%%%%%%%%%%%%%%%%%%%%%%%%%%%%%%%%%%%%%%%%%%%%%%%%%%%%%%%%%%%%%%%%%%%

{\bf Abstract}

\end{center}

\begin{quotation}
  {\small We compute explicitly the first-order in $\alpha'$ corrections to a
    family of solutions of the Heterotic Superstring effective
    action that describes fundamental strings with momentum along themselves,
    parallel to solitonic 5-branes with Kaluza-Klein monopoles
    (Gibbons-Hawking metrics) in their transverse space. These solutions correspond to 4-charge extremal black holes in 4 dimensions upon dimensional
    reduction on $\mathrm{T}^{6}$. We show that some of the $\alpha'$
    corrections can be cancelled by introducing solitonic
    $\mathrm{SU}(2)\times \mathrm{SU}(2)$ Yang-Mills fields, and that this family of
    $\alpha'$-corrected solutions is invariant under $\alpha'$-corrected
    T-duality transformations. We study in detail the
    mechanism that allows us to compute explicitly these $\alpha'$ corrections
    for the ansatz considered here, based on a generalization of the 't~Hooft
    ansatz to hyperK\"ahler spaces.  }
\end{quotation}

\newpage
%%%%%%%%%%%%%%%%%%%%%%%%%%%%%%%%%%%%%%%%%%%%%%%%%%%%%%%%%%%%%%%%%%%%%%
%%%%%%%%%%%%%%%%%%%%%%%%%%%%%%%%%%%%%%%%%%%%%%%%%%%%%%%%%%%%%%%%%%%%%%
%%%%%%%%%%%%%%%%%%%%%%%%%%%%%%%%%%%%%%%%%%%%%%%%%%%%%%%%%%%%%%%%%%%%%%
%%%%%%%%%%%%%%%%%%%%%%%%%%%%%%%%%%%%%%%%%%%%%%%%%%%%%%%%%%%%%%%%%%%%%%
\pagestyle{plain}
%%%%%%%%%%%%%%%%%%%%%%%%%%%%%%%%%%%%%%%%%%%%%%%%%%%%%%%%%%%%%%%%%%%%%%
%%%%%%%%%%%%%%%%%%%%%%%%%%%%%%%%%%%%%%%%%%%%%%%%%%%%%%%%%%%%%%%%%%%%%%
%%%%%%%%%%%%%%%%%%%%%%%%%%%%%%%%%%%%%%%%%%%%%%%%%%%%%%%%%%%%%%%%%%%%%%
%%%%%%%%%%%%%%%%%%%%%%%%%%%%%%%%%%%%%%%%%%%%%%%%%%%%%%%%%%%%%%%%%%%%%%

\tableofcontents

%\newpage

%%%%%%%%%%%%%%%%%%%%%%%%%%%%%%%%%%%%%%%%%%%%%%%%%%%%%%%%%%%%%%%%%%%%%%
%%%%%%%%%%%%%%%%%%%%%%%%%%%%%%%%%%%%%%%%%%%%%%%%%%%%%%%%%%%%%%%%%%%%%%
%%%%%%%%%%%%%%%%%%%%%%%%%%%%%%%%%%%%%%%%%%%%%%%%%%%%%%%%%%%%%%%%%%%%%%
%%%%%%%%%%%%%%%%%%%%%%%%%%%%%%%%%%%%%%%%%%%%%%%%%%%%%%%%%%%%%%%%%%%%%%
\section*{Introduction}
%%%%%%%%%%%%%%%%%%%%%%%%%%%%%%%%%%%%%%%%%%%%%%%%%%%%%%%%%%%%%%%%%%%%%%
%%%%%%%%%%%%%%%%%%%%%%%%%%%%%%%%%%%%%%%%%%%%%%%%%%%%%%%%%%%%%%%%%%%%%%
%%%%%%%%%%%%%%%%%%%%%%%%%%%%%%%%%%%%%%%%%%%%%%%%%%%%%%%%%%%%%%%%%%%%%%
%%%%%%%%%%%%%%%%%%%%%%%%%%%%%%%%%%%%%%%%%%%%%%%%%%%%%%%%%%%%%%%%%%%%%%

Although all the supersymmetric solutions of the Heterotic Superstring
effective action have been classified in Refs.~\cite{Gran:2005wf,Gran:2007fu},
there are many interesting particular solutions yet to be constructed in
detail and studied.

Typically, the construction of the solutions of this theory is made using an
ansatz for $H$, the 3-form field strength of the Kalb-Ramond 2-form $B$, and
its Bianchi identity has to be solved together with the equations of motion of
all the fields. The preferred way of doing this at first order in $\alpha'$ is
to use the analogue of the Green-Schwarz anomaly-cancellation mechanism and
choose a gauge field strength $F$ such that

\begin{equation}
\alpha' \mathrm{Tr}\left[ F\wedge F +R_{(-)}\wedge R_{(-)}\right]=0\, ,   
\end{equation}

\noindent
where $R_{(-)}$ is the curvature 2-form of the torsionful spin connection
$\Omega_{(-)}$ (See {\em e.g.\/} sec.~(\ref{sec-heteroticalpha})). Then, solving the Bianchi identity 

\begin{equation}
  d{H}  
  -
  2\alpha' \mathrm{Tr}\left[ F\wedge F +R_{(-)}\wedge R_{(-)}\right]
  =
  0\, ,
\end{equation}

\noindent
reduces to the much simpler problem of finding a closed 3-form $H$.

This mechanism constrains the gauge field to be essentially identical to, at
least, certain components of the torsionful spin connection. Thus, one may
wish to relax as much as possible this condition so that the gauge field can
have other values or even not be present at all. However, except in some
simple cases, it was not known how to solve the Bianchi identity without using
this mechanism.

In Ref.~\cite{Cano:2018qev} we %have 
observed that in certain cases the
instanton number density $ \mathrm{Tr}\, F\wedge F$ takes the form of the
Laplacian of a function in $\mathbb{E}^{4}$ times the volume 4-form.
Therefore, if $H$ is assumed to be of the form $H\sim
\star_{(4)}d\mathcal{Z}_{0}$ (up to a closed 3-form on $\mathbb{E}^{4}$) for
some function $\mathcal{Z}_{0}$ defined on the same space, the first two terms
in the above Bianchi identity become the Laplacian of a linear combination of
functions with constant coefficients. Almost magically, the third term turns
out to be another Laplacian over the same space and the Bianchi identity is
solved by equating the argument of the Laplacian to zero, up to a harmonic
function on $\mathbb{E}^{4}$. In the case considered in
Ref.~\cite{Cano:2018qev} it was possible to choose the gauge field (a BPST
instanton) so as to achieve the above cancellation, but this was not
completely necessary and one could study the first-order $\alpha'$
corrections to the solution consisting in the harmonic function alone.

The configuration considered in Ref.~\cite{Cano:2018qev} corresponds, after
compactification on $\mathrm{T}^{5}$, to a single, spherically symmetric,
3-charge, extremal 5-dimensional black hole.\footnote{On top
  of the function $\mathcal{Z}_{0}$, its fields are described with another two
  functions, $\mathcal{Z}_{+}$ and $\mathcal{Z}_{-}$.} The modification in the
zeroth-order solution introduced by the gauge field was already known from
non-Abelian gauged 5-dimensional supergravity
\cite{Bellorin:2007yp,Meessen:2015enl,Cano:2017qrq}. The torsionful spin
connection behaves as just another gauge field and, quite remarkably, its
contribution to the $\alpha'$ corrections had to be similar to that of the
instanton, at least in the above Bianchi identity.

From experience, the simplest generalization one can make to this kind of
solutions is to extend the ansatz to multicenter solutions, allowing the
functions occurring in the metric to be arbitrary functions of the
$\mathbb{E}^{4}$ coordinates. In the case of the gauge field, this requires
the use of the so-called \textit{'t~Hooft ansatz} that can describe many BPST
instantons, and is reviewed and generalized in
Appendix~\ref{sec-thooftansatz}. Perhaps not so surprisingly, allowing the
function $\mathcal{Z}_{0}$ to have arbitrary dependence on the
$\mathbb{E}^{4}$ coordinates automatically forces some components of the
torsionful spin connection to take the form of the 't~Hooft ansatz. Then, one
can show that the instanton density 4-forms are, once again, Laplacians, and
the Bianchi identity can be solved in exactly the same way.

It is natural to wonder if this result can be extended further. An interesting
generalization is obtained by replacing $\mathbb{E}^{4}$ with a 4-dimensional
hyperK\"ahler space that has a curvature with the same selfduality properties as
the gauge field. It is well known that the simplest 4-dimensional black holes
one can construct in Heterotic Superstring theory include a Kaluza-Klein
monopole, which is one of the simplest hyperK\"ahler spaces with one
triholomorphic isometry (a Gibbons-Hawking space
\cite{Gibbons:1979zt,Gibbons:1987sp}). The additional isometry is necessary to
obtain a 4-dimensional solution by compactification on
$\mathrm{T}^{6}$. Therefore, this generalization could be used to compute
$\alpha'$ corrections to 4-dimensional black holes such as those considered in
Ref.~\cite{Meessen:2017rwm, Bueno:2014mea}, which also contain non-Abelian
gauge fields.

First of all, one needs to generalize the 't~Hooft ansatz to an arbitrary
hyperK\"ahler space and show that, again, one gets the Laplacian of some
function in that space. We have done this in
Appendix~\ref{sec-thooftansatz}. Now, from the torsionful spin connection we
get terms with the form of this ansatz, which lead to the same result, and
other terms corresponding to the spin connection of the 4-dimensional
hyperK\"ahler manifold. Fortunately, the self-duality properties of these two
contributions are opposite and they do to not mix. However, the contribution
of the latter to the instanton number density might not necessarily take the
form of the Laplacian of some function.

At this stage one could try to add a second $\mathrm{SU}(2)$ gauge field whose
instanton number density cancels that of the hyperK\"ahler manifold. This is
the standard use of the anomaly-cancellation mechanism and has been used, for
this kind of solutions\footnote{Without the additional two functions
  that our class of solutions contains.} in
Ref.~\cite{Papadopoulos:2008rx}. However, it turns out that, if we restrict
ourselves to Gibbons-Hawking spaces, the connection can also be written in an
't~Hooft ansatz-like form that we have called \textit{twisted 't~Hooft ansatz}
(see Appendix~\ref{sec-twistedthooftansatz}) and we get, yet once again, a
combination of Laplacians. Adding a second $\mathrm{SU}(2)$ gauge field is
optional but convenient if we want to cancel the $\alpha'$ corrections.

Thus, for the ansatz we are going to make, we are able to solve the Bianchi
identity of $H$ without invoking the anomaly-cancellation mechanism.

It is somewhat surprising that the equations of motion can be solved as well
in these conditions and there may be another interesting explanation for
it. At any rate, the class of solutions that we find includes all the static, extremal,
(supersymmetric) 4-dimensional black holes of Heterotic Superstring theory and
their first-order in $\alpha'$ corrections, a result that deserves to be
studied and exploited in more detail elsewhere \cite{kn:CChMORR}. In this work
we will just obtain the general solution and we will explain, to the best of
our knowledge, why it can be obtained at all.

%%%%%%%%%%%%%%%%%%%%%%%%%%%%%%%%%%%%%%%%%%%%%%%%%%%%%%%%%%%%%%%%%%%%%%
%%%%%%%%%%%%%%%%%%%%%%%%%%%%%%%%%%%%%%%%%%%%%%%%%%%%%%%%%%%%%%%%%%%%%%
%%%%%%%%%%%%%%%%%%%%%%%%%%%%%%%%%%%%%%%%%%%%%%%%%%%%%%%%%%%%%%%%%%%%%%
%%%%%%%%%%%%%%%%%%%%%%%%%%%%%%%%%%%%%%%%%%%%%%%%%%%%%%%%%%%%%%%%%%%%%%
\subsection*{Self-dual connections and the Atiyah-Hitchin-Singer theorem}
%%%%%%%%%%%%%%%%%%%%%%%%%%%%%%%%%%%%%%%%%%%%%%%%%%%%%%%%%%%%%%%%%%%%%%
%%%%%%%%%%%%%%%%%%%%%%%%%%%%%%%%%%%%%%%%%%%%%%%%%%%%%%%%%%%%%%%%%%%%%%
%%%%%%%%%%%%%%%%%%%%%%%%%%%%%%%%%%%%%%%%%%%%%%%%%%%%%%%%%%%%%%%%%%%%%%
%%%%%%%%%%%%%%%%%%%%%%%%%%%%%%%%%%%%%%%%%%%%%%%%%%%%%%%%%%%%%%%%%%%%%%

Before closing this introduction, it is amusing to think about the relation
between the 't~Hooft ansatz that we use for the Yang-Mills fields and which
arises in the torsionful spin connection and the Atiyah-Hitchin-Singer theorem
Ref.~\cite{Atiyah:1978wi} on self-duality in Riemannian geometry.\footnote{The
  theorem is reviewed and applied to the construction of self-dual Yang-Mills
  instantons on Gibbons-Hawking spaces in \cite{Etesi:2001fb,Etesi:2002cc}.}
The theorem deals with 4-dimensional Riemannian manifolds and the
decomposition of the components of their Levi-Civita spin connection 1-forms
into self- and anti-self-dual combinations according to the well-known local
isomorphism $\mathfrak{so}(4) \cong \mathfrak{su}(2)^{+}\times
\mathfrak{su}(2)^{-}$. We will denote the two terms corresponding to this
decomposition by $\omega^{+\, mn}$, respectively $\omega^{-\, mn}$.  On
the one hand, the theorem states about $\omega^{+\, mn}$ that

\begin{quote}
The curvature 2-form of $\omega^{+\, mn}$ is self-dual if and only if the
manifold is Ricci flat.  
\end{quote}

\noindent
This statement applies, in particular, to hyperK\"ahler manifolds, which are
Ricci flat and, therefore, for them, $\omega^{+\, mn}$ has self-dual
curvature. Moreover, since these have special holonomy $\mathrm{SU}(2)$,
$\omega^{-\, mn}=0$.

On the other hand, the theorem also says that

\begin{quote}
  The curvature 2-form of $\omega^{-\, mn}$ is self-dual if and only if the
  Ricci scalar vanishes and the manifold is conformal to another one with
  self-dual curvature 2-form.
\end{quote}

This can be used to construct self-dual $\mathrm{SU}(2)$ instantons:
consider the metric

\begin{equation}
ds^{2} = P^{2} d\sigma^{2}\, ,  
\end{equation}

\noindent
where $ d\sigma^{2}$ is a hyperK\"ahler metric and where $P$ is some function
defined on it. The Ricci scalar of the full metric is proportional to the
Laplacian of $P$ in the hyperK\"ahler space and vanishes if $P$ is harmonic on
the hyperK\"ahler metric, so in this case the second part of the theorem
applies. If we choose the Vierbein basis $e^{m}=Pv^{m}$ where $v^{m}$ is a
Vierbein basis of the hyperK\"ahler manifold, the first Cartan structure
equation $de^{m}+\omega^{mn}\wedge e^{n}=0$ leads to

\begin{equation}
d\log{P} \wedge v^{m} - \varpi^{mn}\wedge v^{n} +\omega^{mn}\wedge v^{n}=0\, ,
\,\,\,\,\,
\Rightarrow
\,\,\,\,\,
\omega^{mn} = \varpi^{mn} -\partial_{[m}\log{P}\delta_{n]p}v^{p}\, .
\end{equation}

\noindent
where we have used the same equation for the hyperK\"ahler spin connection
$dv^{m}+\varpi^{mn}\wedge v^{n}=0$. We can now project the above equation onto
the anti-self-dual part of $\mathfrak{so}(4)$, i.e. $\mathfrak{su}(2)^{-}$, with
the matrices $(\mathbb{M}^{-}_{mn})^{pq}$ defined in
Eq.~(\ref{eq:M+-matrices}),

\begin{equation}
\omega^{-\, pq} = (\mathbb{M}^{-}_{nm})^{pq}\partial_{m}\log{P} v^{n}\, ,  
\end{equation}

\noindent
and, then, the theorem tells us that the expression in the r.h.s.~is a
connection with self-dual curvature 2-form, or, equivalently, a
$\mathrm{SU}(2)$ gauge connection with self-dual field strength, {\em i.e.\/} an instanton
connection. We prove this fact explicitly in
Appendix~\ref{sec-thooftansatz}. This provides a justification for the
generalized 't~Hooft ansatz that we are using, albeit it does not let one
suspect that the instanton number density will be proportional to a Laplacian.

On the other hand, if we consider the part of the 10-dimensional
metric ansatz conformal to the 4-dimensional hyperK\"ahler manifold, which reads 

\begin{equation}
ds^{2} = \mathcal{Z}_{0} d\sigma^{2}\, ,  
\end{equation}

\noindent
where, at zeroth-order in $\alpha'$, $\mathcal{Z}_{0}$ is a harmonic function
in the hyperK\"ahler manifold. Now the Ricci scalar does not vanish, because
there is a missing factor of $2$ in the exponent of $\mathcal{Z}_{0}$, and the
theorem does not apply. This is, nevertheless, the metric associated to
solitonic 5-branes, and we cannot change it at will. If we repeat the above
calculation we get

\begin{equation}
\omega^{-\, pq} 
= 
\tfrac{1}{2}(\mathbb{M}^{-}_{nm})^{pq}
\partial_{m}\log{\mathcal{Z}_{0}} v^{n}\, ,  
\end{equation}

\noindent
but now the curvature 2-form of this connection will not be
self-dual. Moreover, $\omega^{+\, pq}$ contains the spin connection of the
hyperK\"ahler manifold $\varpi^{mn}$ and some additional terms, which spoil
self-duality in the $\mathfrak{su}(2)^+$ part as well.

This is where the magic of the Heterotic Superstring comes to our rescue
because, now, the object of interest is not the Levi-Civita connection, but
the torsionful spin connection 1-form $\Omega_{(-)}{}^{mn}\equiv
\omega^{mn}-\tfrac{1}{2}H_{p}{}^{mn}e^{p}$, and the contribution of the
torsion is such that  

\begin{equation}
\Omega_{(-)}^{-\, pq} 
=
(\mathbb{M}^{-}_{nm})^{pq}
\partial_{m}\log{\mathcal{Z}_{0}} v^{n}\, ,   
\hspace{2cm}
\Omega_{(-)}^{+\, pq} 
=
\varpi^{mn} \, .
\end{equation}

\noindent
Then, $\Omega_{(-)}^{-\, pq}$ and $\varpi^{mn}$ are both Yang-Mills self-dual
instantons. The curvature 2-form of these connections will, therefore, be
automatically self-dual.

Therefore, in this kind of Heterotic Superstring configurations, the same kind
of objects come up naturally in both the Yang-Mills and in the torsionful spin
connection, via the Atiyah-Hitchin-Singer theorem or via a different
construction which, perhaps, can be related to a generalization of that
theorem. An interesting recent result from
Ref.~\cite{Garcia-Fernandez:2018emx}, which considers the case of compact
spaces, sheds light on this direction. It states that given two instantons on
a given background that satisfies the equations of motion of the heterotic
theory at zeroth order in $\alpha'$, it is always possible to rescale this
background to obtain a solution of first order in $\alpha'$.

The rest of the paper is organized as follows: in
Section~\ref{sec-heteroticalpha} we give a quick review of the low-energy
field theory effective action of the Heterotic Superstring in order to set up the
problem and fix conventions. In Section~\ref{sec-ansatz} we introduce the
ansatz we will work with, although the details of the (generalized) 't~Hooft
ansatz for the gauge fields and its relation with the spin connection of
Gibbons-Hawking spaces are to be found in the Appendices. In
Section~\ref{sec-susy} we show that all the field configurations corresponding
to our ansatz preserve $1/4$ of the 16 possible supersymmetries,
irrespectively of whether they solve the equations of motion or not. 
In Section~\ref{sec-solution} we plug the ansatz into and solve the equations of
motion to first-order in $\alpha'$, using the above mechanism and which is
explained in more detail in the Appendices. In Section~\ref{sec-Tduality} we
study the behavior of the solution under $\alpha'$-corrected T-duality
transformations in the direction in which the strings lie and the waves
propagate (thereby interchanging them), as well as in the isometric direction of the
Gibbons-Hawking space. Finally, in Section~\ref{sec-validity} we make some
general considerations on the validity of these solutions to higher orders in
$\alpha'$.

%%%%%%%%%%%%%%%%%%%%%%%%%%%%%%%%%%%%%%%%%%%%%%%%%%%%%%%%%%%%%%%%%%%%%%
%%%%%%%%%%%%%%%%%%%%%%%%%%%%%%%%%%%%%%%%%%%%%%%%%%%%%%%%%%%%%%%%%%%%%%
%%%%%%%%%%%%%%%%%%%%%%%%%%%%%%%%%%%%%%%%%%%%%%%%%%%%%%%%%%%%%%%%%%%%%%
%%%%%%%%%%%%%%%%%%%%%%%%%%%%%%%%%%%%%%%%%%%%%%%%%%%%%%%%%%%%%%%%%%%%%%
\section{The Heterotic Superstring effective action to
  $\mathcal{O}(\alpha')$}
\label{sec-heteroticalpha}
%%%%%%%%%%%%%%%%%%%%%%%%%%%%%%%%%%%%%%%%%%%%%%%%%%%%%%%%%%%%%%%%%%%%%%
%%%%%%%%%%%%%%%%%%%%%%%%%%%%%%%%%%%%%%%%%%%%%%%%%%%%%%%%%%%%%%%%%%%%%%
%%%%%%%%%%%%%%%%%%%%%%%%%%%%%%%%%%%%%%%%%%%%%%%%%%%%%%%%%%%%%%%%%%%%%%
%%%%%%%%%%%%%%%%%%%%%%%%%%%%%%%%%%%%%%%%%%%%%%%%%%%%%%%%%%%%%%%%%%%%%%

In order to describe the Heterotic Superstring effective action to
$\mathcal{O}(\alpha')$ as given in Ref.~\cite{Bergshoeff:1989de} (but in
string frame), we start by defining the zeroth-order 3-form field strength of
the Kalb-Ramond 2-form $B$:

\begin{equation}
H^{(0)} \equiv dB\, ,  
\end{equation}

\noindent
and constructing with it the zeroth-order torsionful spin connections

\begin{equation}
{\Omega}^{(0)}_{(\pm)}{}^{{a}}{}_{{b}} 
=
{\omega}^{{a}}{}_{{b}}
\pm
\tfrac{1}{2}{H}^{(0)}_{{\mu}}{}^{{a}}{}_{{b}}dx^{{\mu}}\, ,
\end{equation}

\noindent
where ${\omega}^{{a}}{}_{{b}}$ is the Levi-Civita spin connection
1-form.\footnote{We follow the conventions of Ref.~\cite{Ortin:2015hya} for
  the spin connection, the curvature and the gamma matrices.}  With them we
define the zeroth-order Lorentz curvature 2-form and Chern-Simons 3-forms

\begin{eqnarray}
{R}^{(0)}_{(\pm)}{}^{{a}}{}_{{b}}
& = & 
d {\Omega}^{(0)}_{(\pm)}{}^{{a}}{}_{{b}}
- {\Omega}^{(0)}_{(\pm)}{}^{{a}}{}_{{c}}
\wedge  
{\Omega}^{(0)}_{(\pm)}{}^{{c}}{}_{{b}}\, ,
\\
& & \nonumber \\
{\omega}^{{\rm L}\, (0)}_{(\pm)}
& = &  
d{\Omega}^{ (0)}_{(\pm)}{}^{{a}}{}_{{b}} \wedge 
{\Omega}^{ (0)}_{(\pm)}{}^{{b}}{}_{{a}} 
-\tfrac{2}{3}
{\Omega}^{ (0)}_{(\pm)}{}^{{a}}{}_{{b}} \wedge 
{\Omega}^{ (0)}_{(\pm)}{}^{{b}}{}_{{c}} \wedge
{\Omega}^{ (0)}_{(\pm)}{}^{{c}}{}_{{a}}\, .  
\end{eqnarray}

Next, we introduce the gauge fields. We will only activate a
$\mathrm{SU}(2)\times \mathrm{SU}(2)$ subgroup and we will denote by
$A^{A_{1,2}}$ ($A_{1,2}=1,2,3$) the components. The gauge field strength and
the Chern-Simons 3-for of each $\mathrm{SU}(2)$ factor are defined by

\begin{eqnarray}
{F}^{A}
& = & 
d{A}^{A}+\tfrac{1}{2}\epsilon^{ABC}{A}^{B}\wedge{A}^{C}\, , 
\\
& & \nonumber \\
{\omega}^{\rm YM}
& = & 
dA^{A}\wedge {A}^{A}
+\tfrac{1}{3}\epsilon^{ABC}{A}^{A}\wedge{A}^{B}\wedge{A}^{C}\, .
\end{eqnarray}

Then, we are ready to define recursively 

\begin{eqnarray}
H^{(1)}
& = & 
d{B}
+2\alpha'\left({\omega}^{\rm YM}
+{\omega}^{{\rm L}\, (0)}_{(-)}\right)\, ,  
\nonumber \\
& & \nonumber \\
{\Omega}^{(1)}_{(\pm)}{}^{{a}}{}_{{b}} 
& = & 
{\omega}^{{a}}{}_{{b}}
\pm
\tfrac{1}{2}{H}^{(1)}_{{\mu}}{}^{{a}}{}_{{b}}dx^{{\mu}}\, ,
\nonumber \\
& & \nonumber \\
{R}^{(1)}_{(\pm)}{}^{{a}}{}_{{b}}
& = & 
d {\Omega}^{(1)}_{(\pm)}{}^{{a}}{}_{{b}}
- {\Omega}^{(1)}_{(\pm)}{}^{{a}}{}_{{c}}
\wedge  
{\Omega}^{(1)}_{(\pm)}{}^{{c}}{}_{{b}}\, ,
\nonumber  \\
& & \nonumber \\
{\omega}^{{\rm L}\, (1)}_{(\pm)}
& = & 
d{\Omega}^{(1)}_{(\pm)}{}^{{a}}{}_{{b}} \wedge 
{\Omega}^{(1)}_{(\pm)}{}^{{b}}{}_{{a}} 
-\tfrac{2}{3}
{\Omega}^{(1)}_{(\pm)}{}^{{a}}{}_{{b}} \wedge 
{\Omega}^{(1)}_{(\pm)}{}^{{b}}{}_{{c}} \wedge
{\Omega}^{(1)}_{(\pm)}{}^{{c}}{}_{{a}}\, .  
\nonumber \\
& & \nonumber \\
H^{(2)}
& = &  
d{B}
+2\alpha'\left({\omega}^{\rm YM}
+{\omega}^{{\rm L}\, (1)}_{(-)}\right)\, ,  
\end{eqnarray}

\noindent
and so on.

In practice only $\Omega^{(0)}_{(\pm)},{R}^{(0)}_{(\pm)}, \omega^{{\rm L}\,
  (0)}_{(\pm)}, H^{(1)}$ will occur to the order we want to work at, but, often,
it is simpler to work with the higher-order objects ignoring the terms of
higher order in $\alpha'$ when necessary. Thus we will suppress the $(n)$
upper indices.

Finally, we define three ``$T$-tensors'' associated to the $\alpha'$
corrections

\begin{equation}
\label{eq:Ttensors}
\begin{array}{rcl}
{T}^{(4)}
& \equiv &
6
\alpha'\left[
{F}^{A}\wedge{F}^{A}
+
{R}_{(-)}{}^{{a}}{}_{{b}}
\wedge
{R}_{(-)}{}^{{b}}{}_{{a}}
\right]\, ,
\\
& & \\ 
{T}^{(2)}{}_{{\mu}{\nu}}
& \equiv &
2\alpha'\left[
{F}^{A}{}_{{\mu}{\rho}}{F}^{A}{}_{{\nu}}{}^{{\rho}} 
+
{R}_{(-)\, {\mu}{\rho}}{}^{{a}}{}_{{b}}
{R}_{(-)\, {\nu}}{}^{{\rho}\,  {b}}{}_{{a}}
\right]\, ,
\\
& & \\    
{T}^{(0)}
& \equiv &
{T}^{(2)\,\mu}{}_{{\mu}}\, .
\\
\end{array}
\end{equation}

In terms of all these objects, the Heterotic Superstring effective action in
the string frame and to first-order in $\alpha'$ can be written as

\begin{equation}
\label{heterotic}
{S}
=
\frac{g_{s}^{2}}{16\pi G_{N}^{(10)}}
\int d^{10}x\sqrt{|{g}|}\, 
e^{-2{\phi}}\, 
\left\{
{R} 
-4(\partial{\phi})^{2}
+\tfrac{1}{2\cdot 3!}{H}^{2}
-\tfrac{1}{2}T^{(0)}
\right\}\, ,
\end{equation}

\noindent
where $G_{N}^{(10)}$ is the 10-dimensional Newton constant, whose precise
value will not concern us here, $\phi$ is the dilaton field and the vacuum
expectation value of $e^{\phi}$ is the Heterotic Superstring coupling constant
$g_{s}$. $R$ is the Ricci scalar of the string-frame metric
$g_{\mu\nu}$. 

The equations of motion are very complicated, but, following Section~3 of
Ref.~\cite{Bergshoeff:1992cw}, we separate the variations with respect to each
field into those corresponding to occurrences via
${\Omega}_{(-)}{}^{{a}}{}_{{b}}$, that we will call \textit{implicit}, and the
rest, that we will call \textit{explicit}:

\begin{eqnarray}
\delta S 
& = &  
\frac{\delta S}{\delta g_{\mu\nu}}\delta g_{\mu\nu}
+\frac{\delta S}{\delta B_{\mu\nu}}\delta B_{\mu\nu}
+\frac{\delta S}{\delta A^{A_{i}}{}_{\mu}}\delta A^{A_{i}}{}_{\mu}
+\frac{\delta S}{\delta \phi} \delta \phi
\nonumber \\
& & \nonumber \\
& = & 
\left.\frac{\delta S}{\delta g_{\mu\nu}}\right|_{\rm exp.}\delta g_{\mu\nu}
+\left.\frac{\delta S}{\delta B_{\mu\nu}}\right|_{\rm exp.}\delta B_{\mu\nu}
+\left.\frac{\delta S}{\delta A^{A_{i}}{}_{\mu}}\right|_{\rm exp.}
\delta A^{A_{i}}{}_{\mu}
+\frac{\delta S}{\delta \phi} \delta \phi
\nonumber \\
& & \nonumber \\
& &
+\frac{\delta S}{ \delta {\Omega}_{(-)}{}^{{a}}{}_{{b}}}
\left(
\frac{\delta {\Omega}_{(-)}{}^{{a}}{}_{{b}}}{\delta g_{\mu\nu}}
+\frac{\delta {\Omega}_{(-)}{}^{{a}}{}_{{b}}}{\delta B_{\mu\nu}} \delta B_{\mu\nu}
+\frac{\delta {\Omega}_{(-)}{}^{{a}}{}_{{b}}}{\delta A^{A_{i}}{}_{\mu}}\delta
A^{A_{i}}{}_{\mu}
\right)\, .
\end{eqnarray}

We can then apply a lemma proven in Ref.~\cite{Bergshoeff:1989de}: $\delta S/\delta
{\Omega}_{(-)}{}^{{a}}{}_{{b}}$ is proportional to $\alpha'$ and to the
zeroth-order equations of motion of $g_{\mu\nu},B_{\mu\nu}$ and $\phi$ plus
terms of higher order in $\alpha'$.

The upshot is that, if we consider field configurations which solve the
zeroth-order equations of motion\footnote{These can be obtained from
  Eqs.~(\ref{eq:eq1})-(\ref{eq:eq4}) by setting $\alpha'=0$. This eliminates
  the Yang-Mills fields, the $T$-tensors and the Chern-Simons terms in $H$.}
up to terms of order $\alpha'$, the contributions to the equations of motion
associated to the implicit variations are at least of second order in
$\alpha'$ and we can safely ignore them here.

If we restrict ourselves to this kind of field configurations, the equations
of motion reduce to 

\begin{eqnarray}
\label{eq:eq1}
R_{\mu\nu} -2\nabla_{\mu}\partial_{\nu}\phi
+\tfrac{1}{4}{H}_{\mu\rho\sigma}{H}_{\nu}{}^{\rho\sigma}
-T^{(2)}{}_{\mu\nu}
& = & 
0\, ,
\\
& & \nonumber \\
\label{eq:eq2}
(\partial \phi)^{2} -\tfrac{1}{2}\nabla^{2}\phi
-\tfrac{1}{4\cdot 3!}{H}^{2}
+\tfrac{1}{8}T^{(0)}
& = &
0\, ,
\\
& & \nonumber \\
\label{eq:eq3}
d\left(e^{-2\phi}\star {H}\right)
& = &
0\, ,
\\
& & \nonumber \\
\label{eq:eq4}
% e^{2\phi}d\left(e^{-2\phi}\star {F}^{A}\right)
% +\epsilon^{ABC}{A}^{B}\wedge \star F^{C}
% +\star {H}\wedge{F}^{A}
% & = & 
% 0\, .
\alpha' e^{2\phi}\mathfrak{D}_{(+)}\left(e^{-2\phi}\star {F}^{A_{i}}\right)
& = & 
0\, ,
\end{eqnarray}

\noindent
where $\mathfrak{D}_{(+)}$ stands for the exterior derivative covariant with
respect to each $\mathrm{SU}(2)$ subgroup and with respect to the torsionful
connection $\Omega_{(+)}$: suppressing the subindices $1,2$ that distinguish
the two subgroups

\begin{equation}
\label{eq:eq5}
e^{2\phi}d\left(e^{-2\phi}\star {F}^{A}\right)
+\epsilon^{ABC}{A}^{B}\wedge \star F^{C}
+\star {H}\wedge{F}^{A}
= 
0\, .   
\end{equation}

If the ansatz is given in terms of the 3-form field strength we will need to
solve the Bianchi identity 

\begin{equation}
\label{eq:BianchiH}
d{H}  
-
\tfrac{1}{3} T^{(4)}
=
0\, ,
\end{equation}

\noindent
as well.

%%%%%%%%%%%%%%%%%%%%%%%%%%%%%%%%%%%%%%%%%%%%%%%%%%%%%%%%%%%%%%%%%%%%%%
%%%%%%%%%%%%%%%%%%%%%%%%%%%%%%%%%%%%%%%%%%%%%%%%%%%%%%%%%%%%%%%%%%%%%%
%%%%%%%%%%%%%%%%%%%%%%%%%%%%%%%%%%%%%%%%%%%%%%%%%%%%%%%%%%%%%%%%%%%%%%
%%%%%%%%%%%%%%%%%%%%%%%%%%%%%%%%%%%%%%%%%%%%%%%%%%%%%%%%%%%%%%%%%%%%%%
\section{The ansatz}
\label{sec-ansatz}
%%%%%%%%%%%%%%%%%%%%%%%%%%%%%%%%%%%%%%%%%%%%%%%%%%%%%%%%%%%%%%%%%%%%%%
%%%%%%%%%%%%%%%%%%%%%%%%%%%%%%%%%%%%%%%%%%%%%%%%%%%%%%%%%%%%%%%%%%%%%%
%%%%%%%%%%%%%%%%%%%%%%%%%%%%%%%%%%%%%%%%%%%%%%%%%%%%%%%%%%%%%%%%%%%%%%
%%%%%%%%%%%%%%%%%%%%%%%%%%%%%%%%%%%%%%%%%%%%%%%%%%%%%%%%%%%%%%%%%%%%%%

It is convenient to describe our ansatz for each field separately, starting
with the metric, which is assumed to take the general form

\begin{equation}
\label{eq:metric}
ds^{2}
=
\frac{2}{\mathcal{Z}_{-}}du
\left[dv-\tfrac{1}{2}\mathcal{Z}_{+} du\right]
-\mathcal{Z}_{0} d\sigma^{2}
-dy^{i}dy^{i}\, ,
\end{equation}

\noindent
where 

\begin{equation}
\label{eq:metricHK}
d\sigma^{2}=h_{\underline{m}\underline{n}}dx^{m}dx^{n}\, ,
\,\,\,\,\
 m,n=\sharp,1,2,3\, ,
\end{equation}

\noindent
is the metric of a 4-dimensional hyperK\"ahler space and
$\mathcal{Z}_{+},\mathcal{Z}_{-},\mathcal{Z}_{0}$ are functions on
that 4-dimensional space. Thus, the metric is independent of the light-cone
coordinates $u,v$ and of the 4 spatial coordinates $y^{i}$, $i,j=1,2,3,4$. The
hyperK\"ahler metric is characterized by the self-duality of its spin
connection 1-form $\varpi^{mn}$ with respect to the orientation
$\varepsilon^{\sharp 1 2 3 }=+1$ in an appropriate Vierbein basis $v^{m}$

\begin{equation}
\label{eq:HKVierbein}
h_{\underline{m}\underline{n}} = v^{p}{}_{\underline{m}}
v^{p}{}_{\underline{n}}\, .
\end{equation}

\noindent
In order to be able to solve the Bianchi identity of the 3-form $H$ to first
order in $\alpha'$, in Section~\ref{sec-solution} we will find it convenient
to restrict ourselves to Gibbons-Hawking (GH) spaces.

The 3-form field strength is assumed to take the form

\begin{equation}
\label{eq:H}
H 
= 
d\mathcal{Z}^{-1}_{-}\wedge du \wedge dv
+\star_{(4)}d\mathcal{Z}_{0}\, ,  
\end{equation}

\noindent
where $\star_{(4)}$ is the Hodge operator in the 4-dimensional hyperK\"ahler
metric $d\sigma^2$ with the above choice of orientation.

The dilaton field is given by 

\begin{equation}
e^{-2{\phi}}
=
e^{-2{\phi}_{\infty}}\frac{\mathcal{Z}_{-}}{\mathcal{Z}_{0}}\, ,
\end{equation}

\noindent
where $\phi_{\infty}$ is a constant that, in spaces which asymptote to some
vacuum solution, can be identified with the vacuum expectation value,
\textit{i.e.}~$e^{\phi_{\infty}}=g_{s}$.

Finally, we will assume each of the $\mathrm{SU}(2)$ field strengths to live
and be self-dual in the 4-dimensional hyperK\"ahler space with the same
orientation above:

\begin{equation}
F^{A_{1,2}}= +\star_{(4)}F^{A_{1,2}}\, .  
\end{equation}

In order to solve explicitly the equations of motion and, especially, the
Bianchi identity of the 3-form $H$ to first order in $\alpha'$, it is
necessary to know explicitly the 1-form connections. Thus, we are going to
propose an ansatz for them which, as shown in Appendix~\ref{sec-thooftansatz},
automatically gives self-dual 2-form field strengths in hyperK\"ahler
manifolds and which has other advantages that will be discussed later. This
ansatz is most naturally written using pairs of antisymmetric, self-dual,
$\mathrm{SO}(4)$ indices $mn$ as adjoint $\mathrm{SU}(2)$ indices:

\begin{equation}
A_{1,2}^{mn} =(\mathbb{M}^{(-)}_{pq})^{mn}\partial_{q}\log{P_{1,2}}\, v^{p}\, ,
\end{equation}

\noindent
where $\mathbb{M}^{(-)}_{pq}$ are the self-dual generators of
$\mathfrak{so}(4)$, defined in Eq.~(\ref{eq:M+-matrices}),
$\partial_{q}=v_{q}{}^{\underline{m}}\partial_{\underline{m}}$, and the
functions $P_{1}$ and $P_{2}$ are harmonic in the hyperK\"ahler space

\begin{equation}
\nabla^{2}_{(4)}P_{1,2}=0\, .  
\end{equation}

This ansatz generalizes the one recently considered in Ref.~\cite{Cano:2018qev}
for a single, static, 3-charge plus non-Abelian instanton black hole in three
respects: 

\begin{enumerate}
\item No spherical symmetry is assumed: the ansatz can describe multicenter
  configurations.
\item The $\mathbb{R}^{4}$ space transverse to the S5-branes has been replaced
  by an arbitrary hyperK\"ahler space.
\item A second $\mathrm{SU}(2)$ gauge field has been added to the theory. We
  will show that it can be used to suppress $\alpha'$ corrections associated
  to the non-trivial hyperK\"ahler space, just as the first $\mathrm{SU}(2)$
  gauge field can compensate the $\alpha'$ corrections associated to the S5-brane.
\end{enumerate}

%%%%%%%%%%%%%%%%%%%%%%%%%%%%%%%%%%%%%%%%%%%%%%%%%%%%%%%%%%%%%%%%%%%%%%
%%%%%%%%%%%%%%%%%%%%%%%%%%%%%%%%%%%%%%%%%%%%%%%%%%%%%%%%%%%%%%%%%%%%%%
%%%%%%%%%%%%%%%%%%%%%%%%%%%%%%%%%%%%%%%%%%%%%%%%%%%%%%%%%%%%%%%%%%%%%%
%%%%%%%%%%%%%%%%%%%%%%%%%%%%%%%%%%%%%%%%%%%%%%%%%%%%%%%%%%%%%%%%%%%%%%
\section{Supersymmetry of the ansatz}
\label{sec-susy}
%%%%%%%%%%%%%%%%%%%%%%%%%%%%%%%%%%%%%%%%%%%%%%%%%%%%%%%%%%%%%%%%%%%%%%
%%%%%%%%%%%%%%%%%%%%%%%%%%%%%%%%%%%%%%%%%%%%%%%%%%%%%%%%%%%%%%%%%%%%%%
%%%%%%%%%%%%%%%%%%%%%%%%%%%%%%%%%%%%%%%%%%%%%%%%%%%%%%%%%%%%%%%%%%%%%%
%%%%%%%%%%%%%%%%%%%%%%%%%%%%%%%%%%%%%%%%%%%%%%%%%%%%%%%%%%%%%%%%%%%%%%

All the configurations encompassed by our ansatz preserve $1/4$ of the 16
possible supersymmetries, no matter whether they solve the equations of motion
or not. The Killing spinor equations associated to the local supersymmetry
transformations of the gravitino, dilatino and gaugino are, respectively

\begin{eqnarray}
\label{eq:deltapsi}
\nabla_{\mu}^{(+)}\epsilon 
\equiv 
\left(
\partial_{\mu} 
-\tfrac{1}{4}\!\!\not\!\!\Omega_{(+)\, \mu}
\right)\epsilon
& = &
0\, ,
\\
& & \nonumber \\
\label{eq:deltalambda}
\left(
\not\!\partial\phi -\tfrac{1}{12}\!\!\not\!\!H
\right)\epsilon
& = &
0\, ,
\\
& & \nonumber \\
\label{eq:deltachi}
-\tfrac{1}{4}\alpha'\!\!\not\! F^{A_{1,2}} \epsilon
& = &
0\, .
\end{eqnarray}

\noindent
and, using the results of Appendix~\ref{sec-connection} it is easy to see that
the above equations take the same form as in Section~2.1 of
Ref.~\cite{Cano:2018qev}, except for the $m$ components of the first equation,
which receives a contribution from the spin connection of the 4-dimensional
hyperK\"ahler space and the ``doubling'' of the last equation, owed to the
presence of a second $\mathrm{SU}(2)$ gauge field. 

Since the contribution of the spin connection of the 4-dimensional
hyperK\"ahler space is self-dual, just as the contribution coming from the
conformal factor $\mathcal{Z}_{0}$, the $m$ component of the equation
simply gets another term containing the chirality projector
$\tfrac{1}{2}(1-\tilde{\Gamma})$ where $\tilde{\Gamma}\equiv\Gamma^{2345}$ is
the chirality matrix in the 4-dimensional hyperK\"ahler space. Since the two
$\mathrm{SU}(2)$ gauge fields have self-dual field strengths, the two
associated equations (\ref{eq:deltachi}) contain the same chirality projector
$\tfrac{1}{2}(1-\tilde{\Gamma})$ acting on $\epsilon$.

In order to make the paper more self-contained, we write below all the
components of the Killing spinor equations in the frame specified in
Appendix~\ref{sec-connection}

\begin{eqnarray}
\left[
\partial_{+} 
+\tfrac{1}{4}
\frac{\mathcal{Z}_{-}\partial_{m}\mathcal{Z}_{+}}{\mathcal{Z}_{0}^{1/2}}
\Gamma^{m}\Gamma^{+}
\right]\epsilon
& = &
0\, ,
\\
& & \nonumber \\
\left[
\partial_{-} 
+\tfrac{1}{2}\frac{\partial_{m}\log{\mathcal{Z}_{-}}}{\mathcal{Z}_{0}^{1/2}}
\Gamma^{m}\Gamma^{+}
\right]\epsilon
& = &
0\, ,
\\
& & \nonumber \\
\left\{
\partial_{m} 
+\frac{1}{8 \mathcal{Z}_{0}^{1/2}}
\left[
\partial_{q}\log{H}(\mathbb{N}^{+}_{np})_{qm}
+
\partial_{q}\log{\mathcal{Z}_{0}}
(\mathbb{M}^{+}_{qm})_{np}
\right]
\Gamma^{np}(1-\tilde{\Gamma})
\right\}\epsilon
& = &
0\, ,
\\
& & \nonumber \\
\partial_{i} 
\epsilon
& = &
0\, ,
\\
& & \nonumber \\
-\frac{1}{2\mathcal{Z}_{0}^{1/2}}\Gamma^{m}
\left[
\partial_{m}\log{\mathcal{Z}_{-}}\Gamma^{-}\Gamma^{+}
-\partial_{m}\log{\mathcal{Z}_{0}} (1-\tilde{\Gamma})
\right]\epsilon
& = &
0\, ,
\\
& & \nonumber \\
-\frac{1}{8}\alpha'\!\!\not\! F^{A_{1,2}} 
(1-\tilde{\Gamma})\epsilon
& = &
0\, .
\end{eqnarray}

\noindent
We conclude that the Killing spinor equations are solved by constant spinors
satisfying the constraints

\begin{equation}
\tilde{\Gamma}\epsilon=+\epsilon\, ,
\hspace{1cm}
\Gamma^{+}\epsilon=0\, ,
\end{equation}

\noindent
exactly as in the solution studied in Ref.~\cite{Cano:2018qev}.

%%%%%%%%%%%%%%%%%%%%%%%%%%%%%%%%%%%%%%%%%%%%%%%%%%%%%%%%%%%%%%%%%%%%%%
%%%%%%%%%%%%%%%%%%%%%%%%%%%%%%%%%%%%%%%%%%%%%%%%%%%%%%%%%%%%%%%%%%%%%%
%%%%%%%%%%%%%%%%%%%%%%%%%%%%%%%%%%%%%%%%%%%%%%%%%%%%%%%%%%%%%%%%%%%%%%
%%%%%%%%%%%%%%%%%%%%%%%%%%%%%%%%%%%%%%%%%%%%%%%%%%%%%%%%%%%%%%%%%%%%%%
\section{Solving the equations of motion}
\label{sec-solution}
%%%%%%%%%%%%%%%%%%%%%%%%%%%%%%%%%%%%%%%%%%%%%%%%%%%%%%%%%%%%%%%%%%%%%%
%%%%%%%%%%%%%%%%%%%%%%%%%%%%%%%%%%%%%%%%%%%%%%%%%%%%%%%%%%%%%%%%%%%%%%
%%%%%%%%%%%%%%%%%%%%%%%%%%%%%%%%%%%%%%%%%%%%%%%%%%%%%%%%%%%%%%%%%%%%%%
%%%%%%%%%%%%%%%%%%%%%%%%%%%%%%%%%%%%%%%%%%%%%%%%%%%%%%%%%%%%%%%%%%%%%%

Since our ansatz is given in terms of the 3-form field strength, it is
convenient to start by solving its Bianchi identity
Eq.~(\ref{eq:BianchiH}). The fact that it can be solved is one of our
main results.

Due to the structure of our ansatz for $H$, $dH$ is just a Laplacian in the
4-dimensional hyperK\"ahler space:

\begin{equation}
dH= d\star_{(4)}d\mathcal{Z}_{0}
=
-\nabla^{2}_{(4)}\mathcal{Z}_{0}|v| d^{4}x\, .  
\end{equation}

\noindent
The $T^{(4)}$ tensor has three different pieces:

\begin{equation}
{T}^{(4)}
=
6
\alpha'\left[
{F}^{A_{1}}\wedge{F}^{A_{1}}
+
{F}^{A_{2}}\wedge{F}^{A_{2}}
+
{R}_{(-)}{}^{{a}}{}_{{b}}
\wedge
{R}_{(-)}{}^{{b}}{}_{{a}}
\right]\, .
\end{equation}

\noindent
Since we are using a 't~Hooft ansatz for the $\mathrm{SU}(2)$ gauge fields, we
can directly use the result in Eq.~(\ref{eq:FF}). Furthermore, since our
hyperK\"ahler space is, by assumption, a GH space, we can use the result in
Eq.~(\ref{eq:RR}) and, substituting these partial results in
Eq.~(\ref{eq:BianchiH}), we get

\begin{equation}
\nabla^{2}_{(4)}
\left\{
\mathcal{Z}_{0} 
+2\alpha'
\left[ 
(\partial \log{P_{1}})^{2} 
+(\partial \log{P_{2}})^{2}
-(\partial \log{\mathcal{Z}_{0}})^{2} 
-(\partial \log{H})^{2} 
\right]
\right\}
|v| d^{4}x 
 =\mathcal{O}(\alpha'^{2})
\, ,
\end{equation}

\noindent
which is solved exactly to this order by\footnote{The equations are solved
  everywhere except at the singularities of the harmonic function
  $\mathcal{Z}_{0}^{(0)}$, which, in general, will give $\delta$-function
  singularities that, in general, indicate the presence of solitonic
  $5$-branes.    
}

\begin{equation}
\mathcal{Z}_{0} 
=  
\mathcal{Z}^{(0)}_{0}
-2 \alpha' 
\left[ 
(\partial \log{P_{1}})^{2} 
+(\partial \log{P_{2}})^{2}
-(\partial \log{\mathcal{Z}^{(0)}_{0}})^{2} 
-(\partial \log{H})^{2} 
\right]
+\mathcal{O}(\alpha'^{2})\, ,
\end{equation}

\noindent
with

\begin{equation}
\nabla^{2}_{(4)}\mathcal{Z}^{(0)}_{0}=0\, .  
\end{equation}

Some regular gauge fields, when written in the gauge associated to the
't~Hooft anstaz, have singularities that can be removed by a gauge
transformation. However, these unphysical singularities end up contributing to
the instanton number densities $F^{A} \wedge F^{A}$ and $R_{(-)}{}^{a}{}_{b}
\wedge R_{(-)}{}^{b}{}_{a}$ as $\delta$-functions, basically because one is
taking derivatives at points in which the local form of gauge field we are
using becomes singular. In virtue of the removable singularity theorem of
Uhlenbeck Ref.~\cite{Uhlenbeck:1982zm}, it is possible to perform a local
gauge transformation that precisely removes those singularities from the
evaluation of the instanton number densities and, in the preceding expressions
this should carefully be done in the terms inside the squared brackets. Thus,
if the gauge fields are indeed regular, and one has eliminated those
singularies, the only $\delta$-function singularities that remain are those
associated to the harmonic functions $\mathcal{Z}^{(0)}$ and these
singularities will be associated to the presence of branes which source the
fields at the locations of those $\delta$-functions. These delocalized
contributions associated to the instantons correspond, precisely, to the
non-singular terms in brackets.

The removal of the singularities is a very subtle problem, because, in the
end, the hyperK\"ahler space is not part of the physical space, which is the
one that dictates where the physical singularities are and we will not deal
with it here. However, this is an important issue from the physical point of
view which should be discussed in more depth on a case by case basis. We will
make some further comments concerning this point in Section~\ref{sec-Tduality}.

Let us now move to the equations of motion (\ref{eq:eq1})-(\ref{eq:eq4}).

The ansatz automatically solves the Yang-Mills equation
(\ref{eq:eq4})-(\ref{eq:eq5}).

The Kalb-Ramond field equation (\ref{eq:eq3}) reduces to a Laplace equation in
the hyperK\"ahler space

\begin{equation}
\nabla_{(4)}^{2} \mathcal{Z}_{-}
=
0\, .
\end{equation}

Using the expressions above it is straightforward to conclude that the $(++)$
component\footnote{We use the frame specified in equation
  \eqref{eq:framemetric}.} of the Einstein equations (which is the only
non-trivial equation for our ansatz) gives

\begin{equation}
 \mathcal{Z}_{+} = \mathcal{Z}^{(0)}_{+} +\mathcal{O}(\alpha')\, ,
\,\,\,\,\,
\text{with}
\,\,\,\,\,
\nabla_{(4)}^{2} \mathcal{Z}^{(0)}_{+} =0\, ,
\end{equation}

\noindent
with the $\mathcal{O}(\alpha')$ corrections vanishing identically for
Heterotic supergravity. In order to add the stringy corrections one has to
evaluate the $(++)$ component of the $T^{(2)}$ tensor:

\begin{equation}
T^{(2)}_{++}
=
-2\alpha'
R_{(-)\,+abc}R_{(-)\,+}{}^{abc}
=
-2\alpha'\frac{\mathcal{Z}_{-}}{\mathcal{Z}_{0}}
\nabla_{(4)}^{2} 
\left( 
\frac{\partial_{n} \mathcal{Z}^{(0)}_{+}
\partial_{n} \mathcal{Z}_{-}}{\mathcal{Z}^{(0)}_{0}\mathcal{Z}_{-}}
\right)
+
\mathcal{O}(\alpha'^2)\, .
\end{equation}

Then

\begin{equation}
 \mathcal{Z}_{+} 
= 
\mathcal{Z}^{(0)}_{+} 
-4\, \alpha' \left( 
\frac{\partial_{n} \mathcal{Z}^{(0)}_{+}
\partial_{n} \mathcal{Z}_{-}}{\mathcal{Z}^{(0)}_{0}\mathcal{Z}_{-}}
\right)
+
\mathcal{O}(\alpha'^{2})\, .
\end{equation}

\noindent
Obviously, the same comments concerning the removal of spurious singularities
applies to this $\alpha'$ correction.

The dilaton equation (\ref{eq:eq2}) is automatically solved in these
conditions and needs not to be checked explicitly.  Then, given a solution to
$\mathcal{O}(\alpha'^0)$ of the form of our ansatz, which is completely
determined by the harmonic functions $\mathcal{Z}_{+,-,0}^{(0)}$ and
$H^{(0)}$, the most general $\alpha'$-corrected solution of the same form will
be determined by the corrected functions

\begin{eqnarray}
\mathcal{Z}_{+} 
& = & 
\mathcal{Z}_{+}^{(0)}
% +\alpha'\mathcal{Z}_{+}^{(1)} 
-4\, \alpha' \left(
\frac{\partial_{n} \mathcal{Z}_{+}^{(0)}\partial_{n}
  \mathcal{Z}_{-}^{(0)}}{\mathcal{Z}_{0}^{(0)} \mathcal{Z}_-} \right) +\mathcal{O}(\alpha'^{2})\, ,
\\
& & \nonumber \\
 \mathcal{Z}_{-} 
& = & 
\mathcal{Z}_{-}^{(0)}
% +\alpha'\mathcal{Z}_{-}^{(1)}\, ,
+\mathcal{O}(\alpha'^{2})\, ,
\\
& & \nonumber \\
\mathcal{Z}_{0} 
& = & 
\mathcal{Z}_{0}^{(0)}
% +\alpha'\mathcal{Z}_{0}^{(1)}
\nonumber \\
& & \nonumber \\
& & 
-2 \alpha'
\left[
(\partial\log{P_{1}^{(0)}})^{2}
+(\partial\log{P_{2}^{(0)}})^{2} 
-(\partial\log{\mathcal{Z}_{0}^{(0)}})^{2} 
-(\partial\log{H^{(0)}})^{2} \right]
\nonumber \\
& & \nonumber \\
& & 
+\mathcal{O}(\alpha'^{2})
\, ,
\\
& & \nonumber \\
H   
& = &   
H^{(0)}
%+\alpha'  H^{(1)}\, ,
+\mathcal{O}(\alpha'^{2})\, ,
\\
& & \nonumber \\
P_{1,2}  
& = &   
P_{1,2}^{(0)}
% +\alpha'  P_{1,2}^{(1)}\, ,
+\mathcal{O}(\alpha'^{2})\, .
\end{eqnarray}

% \noindent
% where $\mathcal{Z}_{+,-,0}^{(1)}$, $H^{(1)}$ and $P_{1,2}^{(1)}$ are
% completely arbitrary harmonic functions.

This is the main result of our paper. To get a better understanding of this
family of solutions, we are going to study, first, their behavior under
T-duality transformations.

%%%%%%%%%%%%%%%%%%%%%%%%%%%%%%%%%%%%%%%%%%%%%%%%%%%%%%%%%%%%%%%%%%%%%%
%%%%%%%%%%%%%%%%%%%%%%%%%%%%%%%%%%%%%%%%%%%%%%%%%%%%%%%%%%%%%%%%%%%%%%
%%%%%%%%%%%%%%%%%%%%%%%%%%%%%%%%%%%%%%%%%%%%%%%%%%%%%%%%%%%%%%%%%%%%%%
%%%%%%%%%%%%%%%%%%%%%%%%%%%%%%%%%%%%%%%%%%%%%%%%%%%%%%%%%%%%%%%%%%%%%%
\section{$\alpha'$-corrected T-duality}
\label{sec-Tduality}
%%%%%%%%%%%%%%%%%%%%%%%%%%%%%%%%%%%%%%%%%%%%%%%%%%%%%%%%%%%%%%%%%%%%%%
%%%%%%%%%%%%%%%%%%%%%%%%%%%%%%%%%%%%%%%%%%%%%%%%%%%%%%%%%%%%%%%%%%%%%%
%%%%%%%%%%%%%%%%%%%%%%%%%%%%%%%%%%%%%%%%%%%%%%%%%%%%%%%%%%%%%%%%%%%%%%
%%%%%%%%%%%%%%%%%%%%%%%%%%%%%%%%%%%%%%%%%%%%%%%%%%%%%%%%%%%%%%%%%%%%%%

As we have discussed in Section~\ref{sec-ansatz}, the solutions we have found
are a generalization of those studied in Ref.~\cite{Cano:2018qev} with a very
similar structure but more non-trivial harmonic functions that can be
interpreted as describing more extended objects. $\mathcal{Z}_{-,+,0}$,
present in the solution of Ref.~\cite{Cano:2018qev}, are associated,
respectively, to fundamental strings (F1), momentum along the strings (W) and
Neveu-Schwarz (solitonic) 5-branes (S5). $P_{1,2}$ are associated to gauge
5-branes sourced by the instantons. The qualitatively new feature is the
non-trivial hyperK\"ahler space which, generically, describes gravitational
instantons, and the additional (triholomorphic) isometry of this space, which
reduces the possible hyperK\"ahler spaces to be of GH type. These are
completely determined by a harmonic function, $H$. The typical choice
$H=1+1/r$ corresponds to a Kaluza-Klein (KK) monopole, also called (Euclidean)
Taub-NUT space.

In Ref.~\cite{Cano:2018qev} we studied how T-duality acts in the direction of
propagation and winding of the F1 in the presence of first-order $\alpha'$
corrections which affect $\mathcal{Z}_{+}$ but not $\mathcal{Z}_{-}$. At
zeroth order, the standard Buscher rules would simply interchange the complete
$\mathcal{Z}_{+}$ and $\mathcal{Z}_{-}$ functions, including the $\alpha'$
corrections. When first-order corrections are included, this would be wrong
since the dualized solution belongs to the same ansatz and only the
transformed $\mathcal{Z}_{+}'$ can receive $\alpha'$ corrections.

Somewhat extraordinarily, using the $\alpha'$-corrected Buscher rules proposed in
Ref.~\cite{Bergshoeff:1995cg}, we showed that the $\alpha'$ corrections of the
transformed solution only occur where they should and, therefore, the
solutions, as a family, are self-T-dual, as it happens at zeroth-order in
$\alpha'$. This is a highly non-trivial test for both the solutions and the
T-duality rules.

The existence of a second non-trivial isometry in the GH space transverse to
the S5-branes provides us with another non-trivial test. At zeroth order in
$\alpha'$, the single S5-brane solution and the KK monopole are T-dual, and
T-duality simply interchanges their associated harmonic functions
$\mathcal{Z}_{0}$ and $H$. Now, only the former has $\alpha'$ corrections and
T-duality should leave them there. The solutions we have found should be
self-T-dual as a family.

If we perform a T-duality transformation in the direction $x$, the
$\alpha'$-corrected T-duality rules proposed in Ref.~\cite{Bergshoeff:1995cg}
read ($\mu ,\nu\neq \underline{x}$)

\begin{equation}
\begin{array}{rclrcl}
g'_{\mu\nu} & = &
g_{\mu\nu}+\left[ g_{\underline{x}\underline{x}}
G_{\underline{x}\mu}G_{\underline{x}\nu}
-2G_{\underline{x}\underline{x}}G_{\underline{x}(\mu}
g_{\nu)\underline{x}}\right]/G_{\underline{x}\underline{x}}^{2}\, ,
\hspace{-4cm}
\\
& & \\
B'_{\mu\nu} & = &
B_{\mu\nu}-G_{\underline{x}[\mu}
G_{\nu]\underline{x}}/G_{\underline{x}\underline{x}}\, ,
\\
& & \\
g'_{\underline{x}\mu} 
& = & 
-g_{\underline{x}\mu}/G_{\underline{x}\underline{x}}
+g_{\underline{x}\underline{x}}G_{\underline{x}\mu}
/G_{\underline{x}\underline{x}}^{2}\, , 
\hspace{1cm}
&
B'_{\underline{x}\mu} 
& = &
-B_{\underline{x}\mu}/G_{\underline{x}\underline{x}}
-G_{\underline{x}\mu}/G_{\underline{x}\underline{x}}\, ,
\\
& &
\\
g'_{\underline{x}\underline{x}} 
& = &
g_{\underline{x}\underline{x}}/G_{\underline{x}\underline{x}}^{2}\, , 
&
e^{-2\phi'}
& = & 
e^{-2\phi}|G_{\underline{x}\underline{x}}|\, ,
\\
& &
\\
A'^{A}_{\underline{x}} 
& = & 
-A^{A}_{\underline{x}}/G_{\underline{x}\underline{x}}\, , 
&
A'^{A}_{\mu} 
& = &
A^{A}_{\mu} -A^{A}_{\underline{x}}G_{\underline{x}\mu}/G_{\underline{x}\underline{x}}\, ,
\end{array}
\end{equation}

\noindent 
where $G_{\mu\nu}$ (for all the possible values of the indices
$\mu,\nu$ including $\underline{x}$) is
defined by

\begin{equation}
\label{eq:Gmunu}
G_{\mu\nu}
\equiv
g_{\mu\nu}
-B_{\mu\nu}-2\alpha'
\left\{
A^{A}_{\mu}A^{A}{}_{\nu}
+
\Omega_{(-)\, \mu}{}^{a}{}_{b}
{\Omega}_{(-)\, \nu}{}^{b}{}_{a}
\right\}\, .
\end{equation}

The use of these rules requires the explicit knowledge of the components of
the Kalb-Ramond 2-form $B$, which are gauge-dependent. It is natural to use
the gauge of the 't~Hooft ansatz in which the Chern-Simons terms take the
forms computed in Eqs.~(\ref{eq:oYM2}) and (\ref{eq:oL-}), which we reproduce
here for convenience\footnote{According to the discussion in the previous
  section, in certain cases at least, we should eliminate the spurious
  singularities from these Chern-Simons terms. In general, this should simply
  result in a shift by a harmonic function of $\mathcal{Z}_{0}$ that can be
  absorbed in $\mathcal{Z}_{0}^{(0)}$.}

\begin{eqnarray}
\omega^{\rm YM} 
& = &
-\star d
\left[
(\partial \log{P_{1}})^{2}+(\partial \log{P_{2}})^{2}
\right]
+\mathcal{O}(\alpha'^{2})\, ,
\\
& & \nonumber \\
\omega^{\rm L}_{(-)}
& = &
\star_{(4)}d
\left[
(\partial\log{H})^{2}
+
(\partial\log{\mathcal{Z}_{0}})^{2}
\right]
+\mathcal{O}(\alpha'^{2})\, .    
\end{eqnarray}

Then,

\begin{equation}
dB 
= 
H-2\alpha' (\omega^{YM}+\omega^{L}_{(-)})
=
\star_{(4)} d \mathcal{Z}_{0}^{(0)} 
+d\frac{1}{\mathcal{Z}_{-}}\wedge du\wedge dv 
+\mathcal{O}(\alpha'^{2})\, ,
\end{equation}

\noindent
and

\begin{equation}
B
=
\xi_{0}
+\frac{1}{\mathcal{Z}_{-}} du\wedge dv + \mathcal{O}(\alpha'^{2})\,,
\end{equation}

\noindent
where $\xi_{0}=\tfrac{1}{2} \xi_{0\, mn}v^{m}\wedge v^{n}$ is a 2-form on
the hyperK\"ahler space such that 

\begin{equation}
\label{eq:ds=*dZ0}
d\xi_{0}=\star_{(4)} d\mathcal{Z}_{0}^{(0)}\, .
\end{equation}

\noindent
The integrability condition of this equation is the harmonicity of
$\mathcal{Z}_{0}^{(0)}$ in the hyperK\"ahler space, which guarantees the
existence of $\xi_{0}$.

In order to apply the Buscher T-duality rules, one needs to compute the tensor
$G_{\mu\nu}$ defined above in Eq.~(\ref{eq:Gmunu}).  In 10-dimensional flat
indices, its non-vanishing components are\footnote{Observe that some of these
  components have singularities associated to the 't~Hooft ansatz gauge. 
$G_{\mu\nu}$ is not a gauge-invariant quantity, though, and T-duality does not
commute with gauge or Lorentz transformations. Therefore, it is not clear at
all whether these singularities should and can be removed. Again, this is a
problem to be studied on a case by case basis and we will not discuss it here
any further.}

\begin{eqnarray}
G_{++}
& = & 
-4\alpha' \frac{\partial_{m} \mathcal{Z}_{+} \partial_{m}
  \mathcal{Z}_{-}}{\mathcal{Z}_{0}}\, , 
\\ 
& & \nonumber \\
G_{-+}
& = & 
2\, ,
\\
& & \nonumber \\
G_{ij}
& = & 
-\delta_{ij}\, ,
\\ 
& & \nonumber \\
G_{mn}
& = & 
-\delta_{mn}
-\frac{\xi_{0\,    mn}}{\mathcal{Z}_{0}}
-\frac{2\alpha'}{\mathcal{Z}_{0}}
\left\{
\delta_{mn}
\left[
(\partial\log{P_{1}})^{2}
+(\partial\log{P_{2}})^{2}
-(\partial\log{H})^{2}
-(\partial\log{\mathcal{Z}_{0}^{(0)}})^{2}
\right]
\right.
\nonumber \\  
& & \nonumber \\
& & 
-\partial_{m}\log{P_{1}}\partial_{n}\log{P_{1}}
-\partial_{m}\log{P_{2}}\partial_{n}\log{P_{2}}
+\partial_{m}\log{H}\partial_{n}\log{H}
+\partial_{m}\log{\mathcal{Z}_{0}^{(0)}}\partial_{n}\log{\mathcal{Z}_{0}^{(0)}}
\nonumber \\  
& & \nonumber \\
& & 
\left.
+2\partial_{m}\log \mathcal{Z}_{-}\partial_{n}\log \mathcal{Z}_{-} \right\}\, .
\end{eqnarray}

If one makes the coordinate transformation $X=A u+B v$, $Y= Cu + D v$, with $A
D-B C=1$ and then T-dualizes along $X$, we get the T-dual solution

\begin{eqnarray}
d s^{2}{}'
& = & 
\frac{2}{\mathcal{Z}_{-}'}dX\left(dY-\tfrac{1}{2}\mathcal{Z}_{+}' dX\right)
-\mathcal{Z}_{0} h_{\underline{mn}}dx^{m} dx^{n} -dy^{i} dy^{i}\,,
\\
& & \nonumber \\
A^{A_{1,2}}{}'
& = & 
A^{A_{1,2}}\, ,
\\
& & \nonumber \\
B' 
& = & 
\xi_{0}+\left(\frac{B}{D}+\frac{1}{\mathcal{Z}_{-}'}\right) dX\wedge dY \, ,
\\
& & \nonumber \\
e^{2\phi'}
& = & 
e^{2\phi_{\infty}}\frac{\mathcal{Z}_{0}}{\mathcal{Z}_{-}'}\, ,
\end{eqnarray}

\noindent
with

\begin{eqnarray}
\mathcal{Z}_{-}'
& = & 
D\left( 2 C+D \mathcal{Z}^{(0)}_{+} \right)\,,
\\
& & \nonumber \\
\mathcal{Z}_{+}'
& = &
\mathcal{Z}_{-} 
-4 \alpha' \frac{D \partial_{n} \mathcal{Z}_{-}\partial_{n}
   \mathcal{Z}^{(0)}_{+}}{\mathcal{Z}_{0}^{(0)}(2 C+D \mathcal{Z}^{(0)}_{+})}
= 
\mathcal{Z}_{-} 
-4 \alpha' \frac{\partial_{n} \mathcal{Z}_{-}\partial_{n}
  \mathcal{Z}_{-}'}{\mathcal{Z}_{0}^{(0)}\mathcal{Z}_{-}'}\, .
\end{eqnarray}

\noindent
Choosing $C=0, A=D=1$ we preserve the asymptotic behavior of the harmonic
functions and, calling $Y\equiv v'$ and $X\equiv u'$ it is immediate to see that
the T-dual solution belongs to the same family as the original. This is the
result obtained in Ref.~\cite{Cano:2018qev} extended to the presence of a
hyperK\"ahler transverse space.

If $\mathcal{Z}_{+,-,0}, P_{1,2}$ are independent of the coordinate adapted to
the triholomorphic isometry of the GH metric, $z$, (as $H$ is), then the
isometry of the GH space is also an isometry of the full solution and one can
T-dualize it along $z$. In this case, the harmonic functions are harmonic with
respect to 3-dimensional Euclidean space $\mathbb{E}^{3}$ and
Eq.~(\ref{eq:ds=*dZ0}) can be rewritten as

\begin{equation}
d\xi_{0}
=
(dz+\chi)\wedge \star_{(3)} d\mathcal{Z}_{0}^{(0)}
\equiv 
(dz+\chi)\wedge d\chi_{0}\, ,
\,\,\,\,\,
\text{where}
\,\,\,\,\,
\left\{
\begin{array}{rcl}
d\chi & = & \star_{(3)} dH\, ,\\
& & \\
d\chi_{0} & \equiv & \star_{(3)} d\mathcal{Z}_{0}^{(0)}\, ,\\
\end{array}
\right.
\end{equation}

\noindent
This implies that, up to a closed 2-form, 

\begin{equation}
\xi_{0} = \chi_{0}\wedge (dz+\chi)+\tilde{\xi}_{0}\, ,
\end{equation}

\noindent
where $\tilde{\xi}_{0}$ is a 2-form on $\mathbb{E}^{3}$ such that 

\begin{equation}
d \tilde{\xi}_{0} = d\chi\wedge\chi_{0}\, .
\end{equation}

\noindent
Observe that $\tilde{\xi}_{0}$ does not have any $z$ components.

Then, the original solution, written in coordinates adapted to the isometry we
want to T-dualize with respect to, is

\begin{eqnarray}
d s^{2}
& = &
\frac{2}{\mathcal{Z}_{-}}du\left(dv-\tfrac{1}{2}\mathcal{Z}_{+} du\right)
-\mathcal{Z}_{0} \left[\frac{1}{H}(dz+\chi)^{2}+H dx^{r} dx^{r}  \right] 
-dy^{i}dy^{i}\, ,
\\
& & \nonumber \\
A_{1,2}
& = &
\mathbb{M}^{-}_{mn} \partial_{n} \log{P_{1,2}} v^{m} 
 \nonumber \\
& & \nonumber \\
& = &
H^{-1}\mathbb{M}^{-}_{\sharp r} \partial_{\underline{r}}\log{P_{1,2}} (dz+\chi)
 +\mathbb{M}^{-}_{sr} \partial_{\underline{r}}\log{P_{1,2}} dx^{s}\, ,
\\
& & \nonumber \\
B
& = &
\chi_{0}\wedge (dz+\chi)+\tilde{\xi}_{0}
+\frac{1}{\mathcal{Z}_{-}} du\wedge dv \, ,
\\
& & \nonumber \\
e^{2 \phi}
& = &
e^{2\phi_{\infty}}\frac{\mathcal{Z}_{0}}{\mathcal{Z}_{-}}\, .
\end{eqnarray}

\noindent
and the T-dual solution is

\begin{eqnarray}
d s^{2}{}'
& = &
\frac{2}{\mathcal{Z}_{-}}du\left(dv-\tfrac{1}{2}\mathcal{Z}_{+} du\right)
- \mathcal{Z}_{0}' 
\left[
\frac{1}{\mathcal{Z}_{0}^{(0)}}(dz+\chi_{0})^{2}
+\mathcal{Z}_{0}^{(0)} dx^{r} dx^{r}  \right] -dy^{i} dy^{i}\, ,
\\
& & \nonumber \\
A_{1,2}'
& = &
\mathbb{M}^{-}_{mn} \tilde{\partial}_{n} \log{P_{1,2}} \tilde v^{m} 
\nonumber \\
& & \nonumber \\
& = & 
\mathcal{Z}_{0}^{(0)\, -1}\mathbb{M}^{-}_{\sharp r} \partial_{\underline{r}}\log{P} 
(dz+\chi_{0})
+\mathbb{M}^{-}_{sr} \partial_{\underline{r}}\log{P} dx^{s}\, ,
\\
& & \nonumber \\
B'
& = &
\chi_{0}\wedge (dz+\chi)+\tilde{\xi}_{0}'
+\frac{1}{\mathcal{Z}_{-}} du\wedge dv \, ,
\\
& & \nonumber \\
e^{2 \phi}
& = &
e^{2\phi_{\infty}}\frac{\mathcal{Z}_{0}'}{\mathcal{Z}_{-}}\, .
\end{eqnarray}

\noindent
where

\begin{equation}
\mathcal{Z}_{0}'
=
H
-2 \alpha'
\left[ 
(\tilde{\partial}\log{P}_{1})^{2}
+(\tilde{\partial}\log{P}_{2})^{2}
-(\tilde{\partial}\log\mathcal{Z}_{0}^{(0)})^{2} 
-(\tilde{\partial}\log{H})^{2} 
\right]\, ,
\end{equation}

\noindent
$\tilde{\xi}_{0}'$ is a 2-form on $\mathbb{E}^{3}$ defined by  

\begin{equation}
d \tilde{\xi}_{0}' = d\chi_{0}\wedge\chi\, ,
\end{equation}

\noindent
and where $\tilde{\partial}_{m}$ and $\tilde{v}^{m}$ are derivatives in flat
indices and Vierbein associated with the new GH-space obtained substituting
$H\to \mathcal{Z}_{0}^{(0)} $ and, correspondingly $\chi\to \chi_{0}$.

The T-dual solution clearly belongs to the same family as the original and the
net effect of the $\alpha'$-corrected T-duality transformation is the
interchange between the harmonic functions associated to S5-branes and KK
monopoles $\mathcal{Z}_{0}^{(0)}$ and $H$ everywhere, including the $\alpha'$
corrections. This interchange necessarily has to be accompanied by the
interchange of associated 1-forms $\chi_{0}$ and $\chi$. 

This is a highly non-trivial simultaneous test of these $\alpha'$-corrected
solutions and T-duality rules.

%%%%%%%%%%%%%%%%%%%%%%%%%%%%%%%%%%%%%%%%%%%%%%%%%%%%%%%%%%%%%%%%%%%%%%
%%%%%%%%%%%%%%%%%%%%%%%%%%%%%%%%%%%%%%%%%%%%%%%%%%%%%%%%%%%%%%%%%%%%%%
%%%%%%%%%%%%%%%%%%%%%%%%%%%%%%%%%%%%%%%%%%%%%%%%%%%%%%%%%%%%%%%%%%%%%%
%%%%%%%%%%%%%%%%%%%%%%%%%%%%%%%%%%%%%%%%%%%%%%%%%%%%%%%%%%%%%%%%%%%%%%
\section{Range of validity of the solutions}
\label{sec-validity}
%%%%%%%%%%%%%%%%%%%%%%%%%%%%%%%%%%%%%%%%%%%%%%%%%%%%%%%%%%%%%%%%%%%%%%
%%%%%%%%%%%%%%%%%%%%%%%%%%%%%%%%%%%%%%%%%%%%%%%%%%%%%%%%%%%%%%%%%%%%%%
%%%%%%%%%%%%%%%%%%%%%%%%%%%%%%%%%%%%%%%%%%%%%%%%%%%%%%%%%%%%%%%%%%%%%%
%%%%%%%%%%%%%%%%%%%%%%%%%%%%%%%%%%%%%%%%%%%%%%%%%%%%%%%%%%%%%%%%%%%%%%

%Since this is a very wide class of solutions, not much can be said
%about it in full
%generality. 
Since the class of solutions that we are presenting is very wide, not
much can be said  in full detail about the range of validity of the solutions.
It is, however, clear that the same mechanism used in
Ref.~\cite{Cano:2018qev} to cancel the $\alpha'$ corrections in
$\mathcal{Z}_{0}$ can be used here: it is enough to choose $P_{1}=H$ and
$P_{2}=\mathcal{Z}_{0}^{(0)}$ to do it. The result is that all the S5-branes
become symmetric 5-branes. Actually, the solution studied in
Ref.~\cite{Kallosh:1994wy} must be a particular example of this class of
solutions with no $\alpha'$ corrections given by $\mathcal{Z}_{\pm}=1$ and
$\mathcal{Z}_{0}=H$.

The first-order $\alpha'$ corrections in $\mathcal{Z}_{+}$ cannot be cancelled
in the same fashion, at least with the kind of Yang-Mills fields we have used
for our ansatz. The arguments used in Ref.~\cite{Cano:2018qev} suggest that the
second and higher $\alpha'$ corrections can be made arbitrary small or
vanishing if we use the above mechanism to cancel the first-order corrections
of $\mathcal{Z}_{0}$.

%%%%%%%%%%%%%%%%%%%%%%%%%%%%%%%%%%%%%%%%%%%%%%%%%%%%%%%%%%%%%%%%%%%%%%
%%%%%%%%%%%%%%%%%%%%%%%%%%%%%%%%%%%%%%%%%%%%%%%%%%%%%%%%%%%%%%%%%%%%%%
%%%%%%%%%%%%%%%%%%%%%%%%%%%%%%%%%%%%%%%%%%%%%%%%%%%%%%%%%%%%%%%%%%%%%%
%%%%%%%%%%%%%%%%%%%%%%%%%%%%%%%%%%%%%%%%%%%%%%%%%%%%%%%%%%%%%%%%%%%%%%
\section*{Acknowledgments}
%%%%%%%%%%%%%%%%%%%%%%%%%%%%%%%%%%%%%%%%%%%%%%%%%%%%%%%%%%%%%%%%%%%%%%
%%%%%%%%%%%%%%%%%%%%%%%%%%%%%%%%%%%%%%%%%%%%%%%%%%%%%%%%%%%%%%%%%%%%%%
%%%%%%%%%%%%%%%%%%%%%%%%%%%%%%%%%%%%%%%%%%%%%%%%%%%%%%%%%%%%%%%%%%%%%%
%%%%%%%%%%%%%%%%%%%%%%%%%%%%%%%%%%%%%%%%%%%%%%%%%%%%%%%%%%%%%%%%%%%%%%

TO would like to thank Prof.~G. Papadopoulos for pointing us towards
Ref.~\cite{Corrigan:1979di} and him, together with C.~Shahbazi and P.~Cano for
useful conversations.  This work has been supported in part by the
MINECO/FEDER, UE grants FPA2015-66793-P and FPA2015-63667-P, by the Italian
INFN and by the Spanish Research Agency (Agencia Estatal de Investigaci\'on)
through the grant IFT Centro de Excelencia Severo Ochoa SEV-2016-0597. AR is
supported by ``Centro de Excelencia Internacional UAM/CSIC'' and ``Residencia
de Estudiantes'' fellowships. TO wishes to thank M.M.~Fern\'andez for her
permanent support.

%%%%%%%%%%%%%%%%%%%%%%%%%%%%%%%%%%%%%%%%%%%%%%%%%%%%%%%%%%%%%%%%%%%%%%
%%%%%%%%%%%%%%%%%%%%%%%%%%%%%%%%%%%%%%%%%%%%%%%%%%%%%%%%%%%%%%%%%%%%%%
%%%%%%%%%%%%%%%%%%%%%%%%%%%%%%%%%%%%%%%%%%%%%%%%%%%%%%%%%%%%%%%%%%%%%%
%%%%%%%%%%%%%%%%%%%%%%%%%%%%%%%%%%%%%%%%%%%%%%%%%%%%%%%%%%%%%%%%%%%%%%
\appendix
%%%%%%%%%%%%%%%%%%%%%%%%%%%%%%%%%%%%%%%%%%%%%%%%%%%%%%%%%%%%%%%%%%%%%%
%%%%%%%%%%%%%%%%%%%%%%%%%%%%%%%%%%%%%%%%%%%%%%%%%%%%%%%%%%%%%%%%%%%%%%
%%%%%%%%%%%%%%%%%%%%%%%%%%%%%%%%%%%%%%%%%%%%%%%%%%%%%%%%%%%%%%%%%%%%%%

%%%%%%%%%%%%%%%%%%%%%%%%%%%%%%%%%%%%%%%%%%%%%%%%%%%%%%%%%%%%%%%%%%%%%%
%%%%%%%%%%%%%%%%%%%%%%%%%%%%%%%%%%%%%%%%%%%%%%%%%%%%%%%%%%%%%%%%%%%%%%
%%%%%%%%%%%%%%%%%%%%%%%%%%%%%%%%%%%%%%%%%%%%%%%%%%%%%%%%%%%%%%%%%%%%%%
%%%%%%%%%%%%%%%%%%%%%%%%%%%%%%%%%%%%%%%%%%%%%%%%%%%%%%%%%%%%%%%%%%%%%%
\section{Generalized 't~Hooft ansatz in 4d hyperK\"ahler spaces}
\label{sec-thooftansatz}
%%%%%%%%%%%%%%%%%%%%%%%%%%%%%%%%%%%%%%%%%%%%%%%%%%%%%%%%%%%%%%%%%%%%%%
%%%%%%%%%%%%%%%%%%%%%%%%%%%%%%%%%%%%%%%%%%%%%%%%%%%%%%%%%%%%%%%%%%%%%%
%%%%%%%%%%%%%%%%%%%%%%%%%%%%%%%%%%%%%%%%%%%%%%%%%%%%%%%%%%%%%%%%%%%%%%
%%%%%%%%%%%%%%%%%%%%%%%%%%%%%%%%%%%%%%%%%%%%%%%%%%%%%%%%%%%%%%%%%%%%%%

The 6 generators of the Lie algebra $\mathfrak{so}(4)$ in the defining
(vector) representation can be labeled by a pair of antisymmetric indices
$m,n=\sharp,1,2,3$\footnote{Upper and lower indices are identical. The
  positions of the indices are chosen for the sake of clarity.}

\begin{equation}
\label{eq:Mmatrices}
(\mathbb{M}_{mn})^{pq} 
\equiv 
2\delta_{mn}{}^{pq}\, ,
\end{equation}

\noindent
and their commutators are given by

\begin{equation}
[\mathbb{M}_{mn},\mathbb{M}_{pq}]
=
-2\mathbb{M}_{[m|r}(\mathbb{M}_{pq})^{r}{}_{|n]}\, .
\end{equation}

\noindent
These labels are very convenient but they introduce a twofold redundancy, as
each generator appears twice: once as $\mathbb{M}_{\sharp 1}$, for instance, and
once as $\mathbb{M}_{1\sharp}$. Thus, if we want to sum once over all the
independent generators and we sum over these labels, we must introduce
additional factors of $1/2$. For instance, the structure constants have to be
defined by

\begin{equation}
[\mathbb{M}_{mn},\mathbb{M}_{pq}]
\equiv
\tfrac{1}{2}f_{mn\, pq}{}^{rs}\mathbb{M}_{rs}\, ,
\end{equation}

\noindent
and, comparing with the above commutators, we get

\begin{equation}
f_{mn\, pq}{}^{rs} = -4(\mathbb{M}_{pq})^{r}{}_{[m}\delta_{n]}{}^{s}\, . 
\end{equation}

We can define the self- and anti-self-dual combinations 

\begin{equation}
\label{eq:M+-matrices}
% \mathbb{M}_{mn}
% =
% \mathbb{M}^{+}_{mn}
% +
% \mathbb{M}^{-}_{mn}
% \, ,
% \,\,\,\,
\mathbb{M}^{\pm}_{mn}
\equiv 
\tfrac{1}{2}
\left(\mathbb{M}_{mn}\pm
  \tfrac{1}{2}\varepsilon_{mn}{}^{pq}\mathbb{M}_{pq}\right)\, ,
\hspace{1cm}
\tfrac{1}{2}\varepsilon_{mn}{}^{pq}\mathbb{M}^{\pm}{}_{pq}
=
\pm\mathbb{M}^{\pm}{}_{mn}\, ,
\end{equation}

\noindent
which are explicitly given by\footnote{Due to the interchange property, their
  self-duality properties hold in both sets of indices.}

\begin{equation}
(\mathbb{M}^{\pm}_{mn})^{pq}
=
\delta_{mn}{}^{pq}\pm \tfrac{1}{2}\varepsilon_{mn}{}^{pq}
=
(\mathbb{M}^{\pm}_{pq})^{mn}\, ,  
\end{equation}

\noindent
and which must generate two independent subalgebras because they satisfy the
commutation relations

\begin{eqnarray}
[\mathbb{M}^{\pm}_{mn},\mathbb{M}^{\pm}_{pq}]
& = & 
-2\mathbb{M}^{\pm}_{[m|r}(\mathbb{M}^{\pm}_{pq})^{r}{}_{|n]}\, ,
\\
& & \nonumber \\
\left[ \mathbb{M}^{+}_{mn},\mathbb{M}^{-}_{pq} \right] 
& = & 
0\, , 
\end{eqnarray}

\noindent
The (anti-)self-duality properties imply that only three of each kind are
independent and we can pick representatives $\mathbb{M}^{\pm}_{ \sharp i}$, $i=1,2,3$
at the expense of losing manifest $\mathrm{SO}(4)$ covariance. When we work with an
antisymmetric pair of $\mathrm{SO}(4)$ indices, their fourfold redundancy has to be
taken into account introducing factors of $1/4$: 

\begin{equation}
\label{eq:structureconstants+-}
[\mathbb{M}^{\pm}_{mn},\mathbb{M}^{\pm}_{pq}]
\equiv
\tfrac{1}{4}f^{\pm}_{mn\, pq}{}^{rs}\mathbb{M}^{\pm}_{rs}\, ,
\,\,\,\,\,
\Rightarrow
\,\,\,\,\,
f^{\pm}_{mn\, pq}{}^{rs} 
= 
4(\mathbb{M}^{\pm}_{pq})^{x}{}_{[m}(\mathbb{M}^{\pm}_{n]x})^{rs}\, . 
\end{equation}

In order to identify the two 3-dimensional Lie subalgebras, it is convenient
to use the representatives. From the above commutation relations, and with the
convention $\varepsilon_{\sharp 123}=+1$, we find

\begin{equation}
\label{eq:subalgebras}
[\mathbb{M}^{\pm}_{\sharp i},\mathbb{M}^{\pm}_{\sharp j}]
=
\mp \varepsilon_{ijk}\mathbb{M}^{\pm}_{\sharp k}\, . 
\end{equation}

\noindent
Therefore, they are two $\mathfrak{su}(2)$ subalgebras that we are going to
denote by $\mathfrak{su}_{\pm}(2)$. This corresponds to the well known Lie
algebra isomorphism
$\mathfrak{so}(4)=\mathfrak{su}_{+}(2)\oplus\mathfrak{su}_{-}(2)$.

The (anti-)-self-dual combinations can be used in different ways. To start
with, they can be used as a hypercomplex structure in a hyperK\"ahler space in
the basis in which the components are constant.\footnote{This basis may not
  always exist. In that case, one may use the non-constant hypercomplex
  structure to define the ansatz, although some calculations would be more
  complicated to carry out. We thank G.~Papadopoulos for discussions on this
  point.} To fix our conventions and get rid of an excess of $\pm$ and $\mp$
symbols, we are only going to use anti-self-dual hypercomplex structures and
we are going to define

\begin{equation}
J^{i}_{mn} \equiv 2(\mathbb{M}^{-}_{\sharp i})^{mn}\, .
\end{equation}

\noindent
Then, the preservation of the hypercomplex structure by the hyperK\"ahler
space's Levi-Civita connection 1-form $\varpi_{mn}$, 

\begin{equation}
\label{eq:HKdef}
\nabla_{m} J^{i}{}_{np}=0\, , 
\end{equation}

\noindent
implies

\begin{equation}
[\varpi, J^{i}] =0\, ,
\,\,\,\,\,  
\Rightarrow
\,\,\,\,\,\,
\varpi = \varpi^{+}\, ,
\end{equation}

\noindent
so the Levi-Civita connection is self-dual in the $\mathfrak{so}(4)$
indices. The integrability condition of the preservation equation

\begin{equation}
[\nabla_{m},\nabla_{n}] J^{i}{}_{pq} = 0 \, , 
\end{equation}

\noindent
implies

\begin{equation}
[R, J^{i}] =0\, ,
\,\,\,\,\,  
\Rightarrow
\,\,\,\,\,\,
R = R^{+}\, ,
\end{equation}

\noindent
and the Riemann tensor is also self-dual in the $\mathfrak{so}(4)$
indices. This property combined with the Bianchi identity
$\varepsilon^{mnpq}R_{npqr}=0$ leads to one of the main properties of
hyperK\"ahler spaces: their Ricci flatness

\begin{equation}
\label{eq:Ricciflat}
R_{mn}=R_{mpn}{}^{p}=0\, .  
\end{equation}

The second use of the hypercomplex structures we are interested in is the
construction of anti-self-dual $\mathrm{SU}(2)$ instantons through the
so-called \textit{'t~Hooft ansatz}, since they can also be seen as generators
of the $\mathfrak{su}(2)$ algebra. In this context they are usually called
\textit{'t~Hooft symbols} and the following notation is commonly used

\begin{equation}
\eta^{i}{}_{pq} \equiv 2(\mathbb{M}^{(+)}_{\sharp i})^{pq}\, ,
\hspace{1cm} 
\overline{\eta}^{i}{}_{pq} \equiv 2(\mathbb{M}^{(-)}_{\sharp i})^{pq}= J^{i}{}_{pq}\, .
\end{equation}

\noindent
In this case however, we will stick to the $\mathrm{SO}(4)$-covariant
notation, in terms of which the 't~Hooft Ansatz for $\mathrm{SU}(2)$
connection 1-forms reads

\begin{equation}
\label{eq:tHooftansatz}
A^{mn} = (\mathbb{M}^{\pm}_{pq})^{mn}V^{q}v^{p}\, ,   
\end{equation}

\noindent
for some $\mathrm{SO}(4)$ vector field $V^{m}(x)$ and some basis of 1-forms in
the hyperK\"ahler space $v^{m}=v^{m}{}_{\underline{n}}dx^{n}$, related to the
Levi-Civita 1-form connection by 

\begin{equation}
\label{eq:vwconventionHK}
dv^{m}+\varpi^{mn}\wedge v^{n}=0\, ,
\end{equation}

\noindent
in our conventions. In order to compute the corresponding field strength, of
which we will demand self-duality in the spatial indices, we must compute

\begin{equation}
dA = \nabla_{m}\left(\mathbb{M}^{\pm}_{np}V^{p}\right)v^{m}\wedge v^{n}\, ,  
\end{equation}

\noindent
where we are omitting the $\mathrm{SU}(2)\subset \mathrm{SO}(4)$ indices, and,
in order to simplify the computations we are going to assume that

\begin{equation}
\nabla_{m}\mathbb{M}^{\pm}_{np}=0\, ,
\end{equation}

\noindent
where only the lower indices of $\mathbb{M}^{\pm}$ are taken into account in
the covariant derivative.

Thus, except for Euclidean space, whose connection is both self- and
anti-self-dual simultaneously, we can only use one of the two hypercomplex
structures, which will lead to only one kind of instanton field. Since we have
assumed that it is the anti-self-dual hypercomplex structure the one which is
preserved by the connection, we use only that one

\begin{equation}
A = \mathbb{M}^{-}_{mp}V^{p}v^{m}\, .   
\end{equation}

\noindent
With this ansatz, taking into account the commutation relations of the
representatives $ \mathbb{M}^{-}_{0i}$ in Eq.~(\ref{eq:subalgebras}), the
definition for the field strength which leads to the standard $\mathrm{SU}(2)$
Yang-Mills field strength 

\begin{equation}
F^{i}=dA^{i}+\frac{1}{2}\varepsilon^{ijk}A^{j}\wedge A^{k}\, , 
\end{equation}

\noindent
is 

\begin{equation}
F^{mn}= dA^{mn}+A^{mp}\wedge A^{pn}\, ,
\end{equation}

\noindent
and a simple calculation gives

\begin{equation}
F 
= 
-\left\{\tfrac{1}{2} \mathbb{M}^{-}_{mn}V^{p}V^{p} 
+\mathbb{M}^{-}_{mp}(\nabla_{n}V^{p}-V_{n}V^{p})  \right\}v^{m}\wedge v^{n}\, .
\end{equation}

Demanding now self-duality

\begin{equation}
F_{mn}= +\tfrac{1}{2}\varepsilon_{mnpq}F_{pq}\, ,
\,\,\,\,\,
\Rightarrow
\,\,\,\,\,
\nabla_{[m}V_{n]}=0\, ,
\,\,\,\,
\mbox{and}
\,\,\,\,\,  
\nabla_{m}V^{m}+V_{m}V^{m} =0\, ,
\end{equation}

\noindent
which is solved by

\begin{equation}
V_{m} = \partial_{m}\log{P}\, ,
\,\,\,\,\,
\mbox{where}
\,\,\,\,\,
\nabla^{2}P=0\, ,  
\end{equation}

\noindent
so $P$ is a harmonic function on the hyperK\"ahler space. Observe that the
$\mathrm{SU}(2)$ connection and field strengths are both anti-self-dual in the
$\mathrm{SO}(4)$-type gauge indices (which are not shown). However, in the $\mathrm{SO}(4)$
tangent space indices, the field strength is self-dual. There is no chance
that the components $F_{mn}{}^{pq}$ can be interpreted as the components of a
Riemann curvature tensor because, as we have just remarked, $F_{mn}{}^{pq}\neq
F^{pq}{}_{mn}$. We could have made that interpretation if we had demanded
anti-self-duality of the field strength, which leads to more complicated
equations for $V^{m}$.

The $\mathrm{SU}(2)$ Yang-Mills Chern-Simons 3-form, defined in this case
by\footnote{Observe that the trace implies sum over pairs of indices $\sharp
  i, \sharp i$, which can be reexpressed as sums over pairs $mn, nm$ with a
  global minus sign. The latter form allows us to use all the machinery we have
  developed.}

\begin{equation}
\label{eq:oYM}
\omega^{\rm YM}
\equiv 
-\left(dA^{mn}\wedge A^{nm} +\tfrac{2}{3}A^{mn} \wedge A^{np} \wedge A^{pm} \right)\, ,  
\end{equation}

\noindent
takes  for this connection the value

\begin{equation}
\label{eq:oYM2}
\omega^{\rm YM} = -\star dV^{2}= -\star d(\partial \log{P})^{2}\, ,   
\end{equation}

\noindent
where $V^{2}=V^{m}V^{m}$. The instanton number density is, then, given by

\begin{equation}
\label{eq:FF}
F^{A}\wedge F^{A} 
=
d \omega^{\rm YM}
=
-d \star d(\partial \log{P})^{2}  
= 
\nabla^{2} \left[ (\partial \log{P})^{2} \right] |v| d^{4}x\, ,
\end{equation}

\noindent
where $|v|$ is the determinant of the Vierbein or the square root of the
determinant of the metric. In this and other calculations one should be
extremely careful to substract, in the end, any spurious, non-physical
singularities arising from the singularities of the 't~Hooft anstaz, as
explained in Section~\ref{sec-solution}.

The Lorentz Chern-Simons 3-form of a $\mathrm{SO}(4)$ connection
$\Omega^{mn}$ in a 4-dimensional manifold is defined in this case
by\footnote{Observe that now the trace directly implies sum over pairs $mn,
  nm$, which leads to a different global sign.}

\begin{equation}
\omega^{\rm L}
\equiv 
d\Omega^{mn}\wedge \Omega^{nm} +\tfrac{2}{3}\Omega^{mn} \wedge \Omega^{np} \wedge \Omega^{pm} \, .
\end{equation}

\noindent
If the connection $\Omega$ takes the form of the 't~Hooft ansatz in a
hyperK\"ahler space

\begin{equation}
\Omega = \mathbb{M}^{-}_{mp}W^{p}v^{m}\, , 
\hspace{1cm}
W_{m}=\partial_{m}\log{K}\, , 
\,\,\,\,\,
\text{where}
\,\,\,\,\,
\nabla^{2}K=0\, ,
\end{equation}

\noindent
then, 

\begin{equation}
\label{eq:wLK}
\omega^{\rm L}
=
\star dW^{2}= \star d(\partial\log{K})^{2}\, ,  
\end{equation}

\noindent
and

\begin{equation}
R^{mn}\wedge R^{nm} 
=
d \omega^{\rm L}
=
d \star d (\partial\log{K})^{2}
= 
-\nabla^{2} \left[(\partial\log{K})^{2} \right] |v| d^{4}x\, .
\end{equation}

%%%%%%%%%%%%%%%%%%%%%%%%%%%%%%%%%%%%%%%%%%%%%%%%%%%%%%%%%%%%%%%%%%%%%%
%%%%%%%%%%%%%%%%%%%%%%%%%%%%%%%%%%%%%%%%%%%%%%%%%%%%%%%%%%%%%%%%%%%%%%
%%%%%%%%%%%%%%%%%%%%%%%%%%%%%%%%%%%%%%%%%%%%%%%%%%%%%%%%%%%%%%%%%%%%%%
%%%%%%%%%%%%%%%%%%%%%%%%%%%%%%%%%%%%%%%%%%%%%%%%%%%%%%%%%%%%%%%%%%%%%%
\section{The twisted 't~Hooft ansatz in Gibbons-Hawking spaces}
\label{sec-twistedthooftansatz}
%%%%%%%%%%%%%%%%%%%%%%%%%%%%%%%%%%%%%%%%%%%%%%%%%%%%%%%%%%%%%%%%%%%%%%
%%%%%%%%%%%%%%%%%%%%%%%%%%%%%%%%%%%%%%%%%%%%%%%%%%%%%%%%%%%%%%%%%%%%%%
%%%%%%%%%%%%%%%%%%%%%%%%%%%%%%%%%%%%%%%%%%%%%%%%%%%%%%%%%%%%%%%%%%%%%%
%%%%%%%%%%%%%%%%%%%%%%%%%%%%%%%%%%%%%%%%%%%%%%%%%%%%%%%%%%%%%%%%%%%%%%

The metric of hyperK\"ahler spaces admitting a triholomorphic isometry
(Gibbons-Hawking spaces) can always be written in the form \footnote{Here
  $\eta=x^{\sharp}$ and we are using the 3-dimensional, curved, indices
  $x,y,z=1,2,3$ which should not be mistaken with coordinates.}

\begin{equation}
d\sigma^{2}= H^{-1}(d\eta+\chi)^{2}+Hdx^{x}dx^{x}\, ,
\,\,\,\,\,
\partial_{\underline{x}}H
=
\varepsilon_{xyz}\partial_{\underline{y}}\chi_{\underline{z}}\, .  
\end{equation}

In the frame

\begin{equation}
\label{eq:simplestframe}
\begin{array}{rclrcl}
v^{\sharp} & = & H^{-\frac{1}{2}}[d\eta +\chi_{\underline{x}}dx^{x}]\, ,
\hspace{1cm}
& 
v_{\sharp} & = & H^{\frac{1}{2}}\partial_{\eta}\equiv \partial_{\sharp}\, ,\\
& & & & & \\
v^{x} & = & H^{\frac{1}{2}}dx^{x}\, ,
& 
v_{x} & = & H^{-\frac{1}{2}}[\partial_{\underline{x}} 
-\chi_{\underline{x}}\partial_{\eta} ]=\partial_{x}\, ,\\
\end{array}
\end{equation}

\noindent
the non-vanishing components of the Levi-Civita connection
Eq.~(\ref{eq:vwconventionHK}) are given by

\begin{equation}
\label{eq:KKmonopolespinconnection}
\begin{array}{rclrcl}
\varpi_{\sharp\sharp x}
& = & 
-\tfrac{1}{2}\partial_{x} \log{H}\, , 
&
\varpi_{x\sharp y}
& = & 
-\tfrac{1}{2}\epsilon_{xyz}\partial_{z}\log{H}\, ,
\\
& & & & & \\
\varpi_{\sharp xy} 
& = & 
-\tfrac{1}{2}\epsilon_{xyz}\partial_{z}\log{H}\, ,
\hspace{1cm}
&
\varpi_{xyz}
& = & 
\delta_{x[y}\partial_{z]}\log{H}\, ,\\
\end{array}
\end{equation}

\noindent
and they look very similar to those of a $\mathrm{SO}(4)$ connection based on
the 't~Hooft ansatz Eq.~(\ref{eq:tHooftansatz}). As we have explained, the
't~Hooft ansatz does not give a spin connection that can be associated to a
Vierbein, or a proper Riemann tensor and a careful inspection indeed shows
that not all signs of the above components match with that ansatz.

It is possible to \textit{twist} the 't~Hooft ansatz to adapt it to the above
spin connection 1-form, at the expense of breaking the manifest
$\mathrm{SO}(4)$ invariance of the ansatz, which is in agreement with the
existence of an isometric direction in the space. This requires the
introduction of a new set of self- and anti-self-dual $\mathrm{SO}(4)$
generators 

\begin{equation}
\mathbb{N}^{\pm}_{mn} = \pm \tfrac{1}{2}\epsilon_{mnpq}\mathbb{N}^{\pm}_{pq}\, ,
\end{equation}

\noindent
whose representation matrices $(\mathbb{N}^{\pm}_{mn})^{pq}$ have the opposite
self-duality properties, that is

\begin{equation}
(\mathbb{N}^{\pm}_{mn})^{pq} 
= 
\mp\tfrac{1}{2}\epsilon_{pqrs}(\mathbb{N}^{\pm}_{mn})^{rs}\, .
\end{equation}

\noindent
These matrices can be constructed using the $\mathbb{M}^{\pm}_{mn}$ matrices
and a metric $\eta_{mn}=\mathrm{diag}(-+++)$ 

\begin{equation}
(\mathbb{N}^{\pm}_{mn})^{pq} \equiv
\eta_{mr}\eta_{ns}(\mathbb{M}^{\mp}_{rs})^{pq}
\qquad\Rightarrow\qquad 
(\mathbb{N}^{\pm}_{mn})^{pq}=(\mathbb{N}^{\mp}_{pq})^{mn}\, ,
\end{equation}
and satisfy the algebra

\begin{eqnarray}
[\mathbb{N}^{\pm}_{mn},\mathbb{N}^{\pm}_{pq}]
& = & 
-2\mathbb{N}^{\pm}_{[m|r}(\mathbb{N}^{\pm}_{pq})^{st} \eta_{sr}\eta_{t|n]} =-2\mathbb{N}^{\pm}_{[m|r}(\mathbb{M}^{\pm}_{pq})^{r}{}_{|n]}\, ,
\\
& & \nonumber \\
\left[ \mathbb{N}^{+}_{mn},\mathbb{N}^{-}_{pq} \right] 
& = & 
0\, , 
\end{eqnarray}

\noindent
Then, in terms of these matrices, the above spin connection can be rewritten
in the form 

\begin{equation}
\varpi_{mn} = (\mathbb{N}^{+}_{mn})_{pq}\partial_{q}\log{H} v^{p}\equiv (\mathbb{N}^{+}_{mn})_{pq}V^q v^{p}\, ,
\end{equation}
with curvature
\begin{equation}
R^{mn}
= 
-\left\{\tfrac{1}{2} (\mathbb{N}^{+}_{mn})_{rs}V^{p}V^{p} 
+(\mathbb{N}^{+}_{mn})_{rp}(\nabla_{s}V^{p}-V_{s}V^{p})  \right\}v^{r}\wedge v^{s}\, .
\end{equation}

The Chern-Simons 3-form is given by 

\begin{equation}
\label{eq:wLGH}
\omega^{\rm LHK}
=
\star_{(4)}d(\partial\log{H})^{2}\, ,    
\end{equation}

\noindent
and, therefore

\begin{equation}
\label{eq:RR}
R^{mn}\wedge R^{nm}  
=
d\omega^{\rm LHK}
=
d\star_{(4)}d(\partial\log{H})^{2}
=
-\nabla^{2} \left[ (\partial\log{H})^{2} \right] |v|d^4 x\, . 
\end{equation}

%%%%%%%%%%%%%%%%%%%%%%%%%%%%%%%%%%%%%%%%%%%%%%%%%%%%%%%%%%%%%%%%%%%%%%
%%%%%%%%%%%%%%%%%%%%%%%%%%%%%%%%%%%%%%%%%%%%%%%%%%%%%%%%%%%%%%%%%%%%%%
%%%%%%%%%%%%%%%%%%%%%%%%%%%%%%%%%%%%%%%%%%%%%%%%%%%%%%%%%%%%%%%%%%%%%%
%%%%%%%%%%%%%%%%%%%%%%%%%%%%%%%%%%%%%%%%%%%%%%%%%%%%%%%%%%%%%%%%%%%%%%
\section{Connections and curvatures}
\label{sec-connection}
%%%%%%%%%%%%%%%%%%%%%%%%%%%%%%%%%%%%%%%%%%%%%%%%%%%%%%%%%%%%%%%%%%%%%%
%%%%%%%%%%%%%%%%%%%%%%%%%%%%%%%%%%%%%%%%%%%%%%%%%%%%%%%%%%%%%%%%%%%%%%
%%%%%%%%%%%%%%%%%%%%%%%%%%%%%%%%%%%%%%%%%%%%%%%%%%%%%%%%%%%%%%%%%%%%%%
%%%%%%%%%%%%%%%%%%%%%%%%%%%%%%%%%%%%%%%%%%%%%%%%%%%%%%%%%%%%%%%%%%%%%%

In this appendix we are going to compute the Levi-Civita and torsionful spin
connections and their associated Chern-Simons terms and curvatures for our
ansatz, which is described in Section~\ref{sec-ansatz}.

A simple choice of Zehnbein is 

\begin{equation}
\label{eq:framemetric}
e^{+} = \frac{du}{\mathcal{Z}_{-}}\, , 
\quad 
e^{-} = dv-\tfrac{1}{2}\mathcal{Z}_{+} du\, , 
\quad 
e^{m} = \mathcal{Z}_{0}^{1/2}\, v^{m}\, , 
\quad 
e^{i}=dy^{i}\, ,
\end{equation}

\noindent
where $v^{m}=v^{m}{}_{\underline{n}}dx^{n}$ is a Vierbein of the
four-dimensional hyper-K\"ahler space defined in Eq.~(\ref{eq:HKVierbein}).
The inverse basis is 

\begin{equation}
e_{+} 
=
\mathcal{Z}_{-}(\partial_{u}+\tfrac{1}{2}\mathcal{Z}_{+}\partial_{v})\, ,
\quad 
e_{-} 
=
\partial_{v}\, ,
\quad
e_{m}
=
\mathcal{Z}_{0}^{-1/2}\partial_{m}\, ,
\quad
e_{i}=\partial_{i}\, ,
\end{equation}

\noindent
where $\partial_{m}\equiv v_{m}{}^{\underline{n}}\partial_{\underline{n}}$ is
the inverse basis in the hyperK\"ahler space and any other $m,n$ index will be
a flat index in the hyperK\"ahler space and will be raised and lowered with
$+\delta_{mn}$.

Using the structure equation $de^{a}=\omega^{a}{}_{b}\wedge e^{b}$ we find
that the non-vanishing components of the spin connection are given by 

\begin{equation}
\begin{array}{rclrcl}
\omega_{-+m} 
& = & 
\omega_{+-m} 
= 
\omega_{m+-} 
=
{\displaystyle\frac{1}{2 \mathcal{Z}_{0}^{1/2}}
\partial_{m}\log{\mathcal{Z}_{-}}\, ,} 
\quad 
&
\omega_{++m} 
& = & 
{\displaystyle
\frac{\mathcal{Z}_{-}}{2\mathcal{Z}^{1/2}_{0}}\partial_{m}\mathcal{Z}_{+}}\, , 
\\
& & & & & \\
\omega_{mnp} 
& = &
\mathcal{Z}_{0}^{-1/2}
\left[
\varpi_{mnp}
+
\frac{1}{2} (\mathbb{M}_{mq})_{np}\partial_{q}\log{ \mathcal{Z}_{0}}
\right]\, ,
&  &  & 
\end{array}
\end{equation}

\noindent
where $\varpi_{mnp}$ are the components of the spin connection on the
hyperK\"ahler space defined with the convention
Eq.~(\ref{eq:vwconventionHK}).\footnote{These 4-dimensional tangent-space
  indices are raised and lowered with $+\delta_{mn}$ and there is no
  difference between them, beyond an esthetic one.} We assume they satisfy the
properties Eq.~(\ref{eq:HKdef})-(\ref{eq:Ricciflat}) with the conventions we
use.

% In curved indices, the non-vanishing components of the spin connection are

% \begin{equation}
% \begin{array}{rclrcl}
% \omega_{\underline{m}+-} 
% & = &
% \tfrac{1}{2}\partial_{\underline{m}}\log{\mathcal{Z}_{-}}\, ,
% \hspace{1cm}
% &
% \omega_{u+m} 
% & = &
% -\tfrac{1}{4}{\displaystyle
%   \frac{\mathcal{Z}_{+}}{\mathcal{Z}_{0}^{1/2}}\partial_{m}\log{\mathcal{Z}_{-}}}\, ,
% \\    
% & & & & & \\
% \omega_{v+m} 
% & = &
% {\displaystyle\frac{1}{2\mathcal{Z}_{0}^{1/2}}}\partial_{m}\log{\mathcal{Z}_{-}}\, ,
% &
% \omega_{u-m} 
% & = &
% -{\displaystyle\frac{1}{2\mathcal{Z}_{0}^{1/2}}}\partial_{m}\mathcal{Z}_{-}^{-1}\, ,
% \\
% & & & & & \\
% \omega_{\underline{p}mn}
% & = &
% \varpi_{\underline{p}mn}
% +
% v_{p}{}^{\underline{p}}(\mathbb{M}_{pq})_{mn}\partial_{q} \log{\mathcal{Z}_{0}}
% \, ,\hspace{-2cm}
% & & & 
% \\
% \end{array}
% \end{equation}

In order to compute the components of the torsionful spin connections, we need
the components of the 3-form field strength. From Eq.~(\ref{eq:H}), in the
above Zehnbein basis they are given by

\begin{equation}
H_{m+-}
 =  
-
\mathcal{Z}_{0}^{-1/2}\partial_{m}\log{\mathcal{Z}_{-}}\, ,
\hspace{1cm}
H_{mnp}
= 
\mathcal{Z}_{0}^{-1/2} \, 
\varepsilon_{mnpq}\partial_{q}\log{\mathcal{Z}_{0}}\, .
\end{equation}

Then, the non-vanishing flat components of the torsionful spin connection
$\Omega_{(-)abc}\equiv \omega_{abc}-\tfrac{1}{2}H_{abc}$ are

\begin{equation}
\begin{array}{rclrcl}
\Omega_{(-)+-m}
& = &
\Omega_{(-)m+-}
=
\mathcal{Z}_{0}^{-1/2}\partial_{m}\log{\mathcal{Z}_{-}}\, ,
&
\Omega_{(-)++m}
& = &
\tfrac{1}{2}
\mathcal{Z}_{-}\mathcal{Z}_{0}^{-1/2}\partial_{m}\mathcal{Z}_{+}\, ,
\\    
& & & & & \\
\Omega_{(-)mnp}
& = &
\mathcal{Z}_{0}^{-1/2}
\left[
\varpi_{mnp}
+(\mathbb{M}^{-}_{mq})_{np}\partial_{q}\log{\mathcal{Z}_{0}}
\right]\, ,
& & & \\
\end{array}
\end{equation}

\noindent
and those of $\Omega_{(+)abc}\equiv \omega_{abc}+\tfrac{1}{2}H_{abc}$ are
given by 

\begin{equation}
\begin{array}{rclrcl}
\Omega_{(+)-+m}
& = &
\mathcal{Z}_{0}^{-1/2}\partial_{m}\log{\mathcal{Z}_{-}}\, ,
&
\Omega_{(+)++m}
& = &
\tfrac{1}{2}
\mathcal{Z}_{-}\mathcal{Z}_{0}^{-1/2}\partial_{m}\mathcal{Z}_{+}\, ,
\\    
& & & & & \\
\Omega_{(+)mnp}
& = &
\mathcal{Z}_{0}^{-1/2}
\left[
\varpi_{mnp}
+
(\mathbb{M}^{+}_{mq})_{np}\partial_{q}\log{\mathcal{Z}_{0}}
\right]\, ,
& & & \\
\end{array}
\end{equation}

\noindent
where the $4\times 4$ matrices $\mathbb{M}^{\pm}_{np}$ are defined in
Eq.~(\ref{eq:M+-matrices}).

The Lorentz-Chern-Simons 3-form $\omega^{\rm L}_{(-)}$ reduces to the
Chern-Simons 3-form of the $\mathrm{SO}(4)$ connection $\Omega_{(-)mn}$

\begin{equation}
  \begin{array}{rcl}
\omega^{\rm L}_{(-)}  
& \equiv &
d\Omega_{(-)}{}^{{a}}{}_{{b}} \wedge 
\Omega_{(-)}{}^{{b}}{}_{{a}} 
-\tfrac{2}{3}
\Omega_{(-)}{}^{{a}}{}_{{b}} \wedge 
\Omega_{(-)}{}^{{b}}{}_{{c}} \wedge
\Omega_{(-)}{}^{{c}}{}_{{a}}
\\
& & \\
& = & 
d\Omega_{(-)mn} \wedge 
\Omega_{(-)nm} 
+\tfrac{2}{3}
\Omega_{(-)mn} \wedge 
\Omega_{(-)np} \wedge
\Omega_{(-)pm}\, ,
\end{array}
\end{equation}

\noindent
which, in its turn, is just the sum of the Chern-Simons 3-forms of the
self-dual and anti-self-dual pieces of $\Omega_{(-)mn}$: the self-dual spin
connection of the hyperK\"ahler manifold and the anti-self-dual 1-form
$(\mathbb{M}^{-}_{mq})_{np}\partial_{q}\log{\mathcal{Z}_{0}}$. The latter has
the form of the 't~Hooft ansatz Eq.~(\ref{eq:tHooftansatz}) discussed in
Appendix~\ref{sec-thooftansatz} and, therefore, its Chern-Simons term takes
the value computed in Eq.~(\ref{eq:wLK}) with $K$ replaced by
$\mathcal{Z}_{0}$. The Chern-Simons 3-form of the spin connection of the
hyperK\"ahler manifold has to be computed case by case, except when it is a
Gibbons-Hawking space. In that case, there is a general expression for it (See
{\em e.g.\/} Eq.~(\ref{eq:wLGH})) and for its total derivative which are
particularly convenient for us because the Bianchi identity of the 3-form
field strength $H$ becomes a linear combination of Laplacians on the
Gibbons-Hawking space that can be solved exactly.

Then, in these conditions, we have

\begin{equation}
\label{eq:oL-}
\omega^{\rm L}_{(-)}
=
\star_{(4)}d
\left[
(\partial\log{H})^{2}
+
(\partial\log{\mathcal{Z}_{0}})^{2}
\right]\, ,    
\end{equation}

\noindent
and

\begin{equation}
\mathrm{Tr}(R_{(-)}\wedge R_{(-)})  
=
d\omega^{\rm L}_{(-)}
=
%d\star_{(4)}d
%\left[
%(\partial\log{H})^{2}
%+
%(\partial\log{\mathcal{Z}_{0}})^{2}
%\right]
%=
-\nabla^{2}
\left[
(\partial\log{H})^{2}
+
(\partial\log{\mathcal{Z}_{0}})^{2}
\right] \, .
\end{equation}

Clearly, it would be extremely interesting to find other hyperK\"ahler spaces with no
  triholomorphic isometry that still enjoy the same property. The Atiyah-Hitchin hyperK\"ahler space
  \cite{Atiyah:1985dv}, which has been considered before in the context of
  supergravity solutions in Refs.~\cite{Bena:2007ju, Halmagyi:2017lqm}, might
  provide an explicit example. We leave this study for future
  work. Interestingly, for arbitrary self-dual $\mathrm{SU}(2)$ instanton
  fields on $\mathbb{R}^{4}$, and not just for those in the 't~Hooft ansatz,
  this {\it{Laplacian property}} was proven in Ref.~\cite{Corrigan:1979di}
  using the ADHM construction \cite{Drinfeld:1978xr,Atiyah:1978ri}.  Our
  results suggest that this property could also hold in hyperK\"ahler
  backgrounds and, therefore, for the spin connections of the hyperK\"ahler
  spaces themselves, as it happens in Gibbons-Hawking spaces.

%%%%%%%%%%%%%%%%%%%%%%%%%%%%%%%%%%%%%%%%%%%%%%%%%%%%%%%%%%%%%%%%%%%%%%
%%%%%%%%%%%%%%%%%%%%%%%%%%%%%%%%%%%%%%%%%%%%%%%%%%%%%%%%%%%%%%%%%%%%%%
%%%%%%%%%%%%%%%%%%%%%%%%%%%%%%%%%%%%%%%%%%%%%%%%%%%%%%%%%%%%%%%%%%%%%%
%%%%%%%%%%%%%%%%%%%%%%%%%%%%%%%%%%%%%%%%%%%%%%%%%%%%%%%%%%%%%%%%%%%%%%
%%%%%%%%%%%%%%%%%%%%%%%%%%%%%%%%%%%%%%%%%%%%%%%%%%%%%%%%%%%%%%%%%%%%%%
%%%%%%%%%%%%%%%%%%%%%%%%%%%%%%%%%%%%%%%%%%%%%%%%%%%%%%%%%%%%%%%%%%%%%%

\end{document}